\documentclass[suppldata]{interact}
\usepackage{fix-cm}
\usepackage{epstopdf}
\usepackage{graphicx}
\usepackage{float}
\usepackage{footnote}
\usepackage{xfrac}
\usepackage{here}
\usepackage{url}
\usepackage{subcaption} 
\usepackage[margin=10pt,font={sf,small},labelfont=sf]{caption}
\usepackage{color}
\usepackage{natbib}
\usepackage{algorithm}
\usepackage{algpseudocode}
\usepackage{etoolbox}

\theoremstyle{plain}
\newtheorem{theorem}{Theorem}[section]

\theoremstyle{definition}
\newtheorem{definition}[theorem]{Definition}

\theoremstyle{remark}

\newcommand{\be}{\begin{equation}}
\newcommand{\ee}{\end{equation}}

\makeatletter

\newcommand*{\algrule}[1][\algorithmicindent]{
  \makebox[#1][l]{
    \hspace*{.2em}
    \vrule height .75\baselineskip depth .25\baselineskip
  }
}

\newcount\ALG@printindent@tempcnta
\def\ALG@printindent{
    \ifnum \theALG@nested>0
    \ifx\ALG@text\ALG@x@notext
    \else
    \unskip
    \ALG@printindent@tempcnta=1
    \loop
    \algrule[\csname ALG@ind@\the\ALG@printindent@tempcnta\endcsname]
    \advance \ALG@printindent@tempcnta 1
    \ifnum \ALG@printindent@tempcnta<\numexpr\theALG@nested+1\relax
    \repeat
    \fi
    \fi
}
\patchcmd{\ALG@doentity}{\noindent\hskip\ALG@tlm}{\ALG@printindent}{}{\errmessage{failed to patch}}
\patchcmd{\ALG@doentity}{\item[]\nointerlineskip}{}{}{}
\makeatother

\begin{document}

\title{\begin{center} 
Monitoring 3D Lattice Structures in Additive Manufacturing Using Topological Data Analysis \\
 \end{center}}

\author{
\begin{center}
\name{Yulin An\textsuperscript{$\ast$}\thanks{Y.A. e-mail: yba5115@psu.edu}, Xueqi Zhao\textsuperscript{$\dagger$}\thanks{X.Z. e-mail xuz206.psu@gmail.com} and Enrique del Castillo\textsuperscript{$\ast,\ddagger$}\thanks{E.D.C., corresponding author, e-mail: exd13@psu.edu}}
\textsuperscript{$\ast$} Department of Industrial and Manufacturing Engineering\\
The Pennsylvania State University, University Park, PA\\
 \textsuperscript{$\dagger$} Google Cloud, Google LLC., Mountain View, CA\\
 \textsuperscript{$\ddagger$}Department of Statistics\\ The Pennsylvania State University, University Park, PA
\end{center}
}

\maketitle

\begin{abstract}

We present a new method for the statistical process control of lattice structures using tools from Topological Data Analysis. Motivated by applications in additive manufacturing, such as aerospace components and biomedical implants, where hollow lattice geometries are critical, the proposed framework is based on monitoring the persistent homology properties of parts. Specifically, we focus on homological features of dimensions zero and one, corresponding to connected components and one-dimensional loops, to characterize and detect changes in the topology of lattice structures. A nonparametric hypothesis testing procedure and a control charting scheme are introduced to monitor these features during production. Furthermore, we conduct extensive run-length analysis via various simulated but real-life lattice-structured parts. Our results demonstrate that persistent homology is well-suited for detecting topological anomalies in complex geometries and offers a robust, intrinsically geometrical alternative to other SPC methods for mesh and point data.
\end{abstract}


\section{Introduction}
How to optimally monitor and control complex systems has always been one of the most crucial aspects of modern quality engineering. One fundamental type of complexity occurs when data lie on a lower-dimensional, curved subspace or manifold of the ambient space. A canonical example of manifold data occurring in manufacturing is surface measurements produced by 3D non-contact (range) scanners, where discrete parts are produced and measured. In its most unprocessed form, scanner data have the form of unstructured point clouds, i.e., datasets where only the $(x_i, y_i, z_i)$ coordinates of hundreds of thousands of points $\bf p_i$ on the surface of a part are recorded. The scanner is not perfect, so the true surface observed with measurement errors in all 3 coordinates is the object of interest. In addition to the unstructured point cloud data, it is frequent that industrial 3D scanners generate mesh data, and in particular, triangulations, where the measured points correspond to the vertices, linked to nearby 
so that they form a network of triangles that defines the surface of an object. Such triangulations are also common in Computer-Aided Design (CAD) and in Additive Manufacturing (AM), and provide additional information on the topology of the part not conveyed by point clouds.

To identify possible differences between the 3D objects, traditional Shape Analysis methods first superimpose, or register, two datasets representing two objects, so that they are aligned into the same location, orientation, and scale and are therefore comparable \citep{Drydmard16}. Then the distance between the datasets can be evaluated and used to determine whether the two underlying objects are similar enough. Recently, \cite{zhaoEDC_Tech} proposed a registration-free Statistical Process Control (SPC) approach, which, for 3D parts, aims to directly compare \textit{intrinsic} geometrical features, those that do not depend on the ambient space where they are located, of a sequence of scanned part surfaces. By using only the intrinsic geometrical properties of each part, the computationally expensive registration step can be avoided. Besides the computational savings, intrinsic geometrical methods, which do not require registration, are particularly practical for data acquired with portable, hand-held scanners or computed tomography (CT) machines. 

A different family of intrinsic methods to handle geometric data that can have applications in monitoring lattice-like structures in manufacturing is Topological Data Analysis (TDA) methods \citep{Wasserman}. In TDA, a set of points in $\mathbb{R}^n$, i.e., a point cloud, is observed. It is assumed that these points are measurements on some underlying manifold, and the goal is to make statistical inferences on the shape of the manifold from the point cloud data. An important TDA tool, described in Appendix \ref{app:A}, is the persistent homology (PH) diagram of an object.  In this paper, we propose new methods for statistical inference based on the PH diagrams (or {\em persistence diagrams}) of point clouds obtained from part scans in order to perform statistical process control of complex objects in manufacturing, in particular, lattice-like structures that contain a non-trivial topology. PH extracts topological features from a manifold, such as the number of connected components and the number of cycles of different dimensions, based on a simplicial complex that is dynamically changing. Persistence diagrams are a graphical representation of this dynamic exploration of the manifold. We propose to monitor the persistence diagrams to evaluate the quality of a lattice structure in Additive Manufacturing and other manufacturing processes where they are produced. 

The rest of the paper is organized as follows. In the next section, we first review related work on topological data analysis. Extracting topological features via PH is a computational task, and therefore, in Section \ref{Compute} of this paper, we investigate and compare the performance of two R packages that compute persistence diagrams. Next, in Section \ref{TDA.dist}, we introduce different distance measures between the persistence diagrams of two objects. Once equipped with a metric in the space of persistence diagrams, a permutation test is developed in Section \ref{TDA.permtest} to statistically compare the persistence of one object against that of a group of objects deemed to be produced in a state of statistical control. The average run length performance of the permutation test used for nonparametric SPC is studied for a variety of lattice-like objects in Section \ref{TDA.SPC}. Section \ref{TDA.CaseStudy} presents a simulation case study based on a computer-aided design (CAD) model of a real additive manufacturing lattice part. Whenever presenting a new methodology in a field, one must present a comparative study of its performance against existing alternatives. There are not many other existing methods in the literature for monitoring the complete 3D geometry of parts. The best such procedure, and a good comparison with the PH-based method proposed in this paper, as these are both intrinsic geometrical methods, is the spectral SPC methods \citep{an2025practical, zhaoEDC_Tech}.  A comparison between the DFEWMA chart based on the LB spectrum and the PH-based permutation test SPC method of this paper is considered in the last section of this paper, Section \ref{TDAvsLB}.

\section{Related Work}
Topological Data Analysis (TDA), although a recent area of intensive research in Statistics, is based on ideas which appear with the development of Combinatorial Topology methods at the beginning of the 20 century by Poincar\'{e} \citep{buchanan2025lasting}. 
Topologists started working on the concept of {Homology} from the intuition that objects of similar shape have the same number of holes. Early authors tried to provide a classification of different manifolds, with the earliest such attempt being Euler’s characteristic for regular polyhedra, and more recently, the Betti numbers (see Appendix \ref{app:A}). Homology, which quantifies the number of holes or cycles in a manifold, is computed using simplicial complexes \citep{Alexandrov1}.

\cite{Wasserman} reviews some main TDA methods, namely clustering, nonlinear dimension reduction, and persistent homology. 
In an unpublished Ph.D. dissertation work, \cite{zhao2022intrinsic} compared various software packages to calculate the persistent homology (PH) diagrams for point clouds, in particular, the performance of two R implementations,  packages \texttt{TDA} and \texttt{TDAstats}, and concluded that the \texttt{TDAstats} package computes the persistent homology of 0-dimensional and 1-dimensional features quickly, but it struggles whenever higher-dimensional features are of interest. On the other hand, although the \texttt{TDA} package generally computes PH diagrams more slowly, its performance is better when computing higher-dimensional cycles \citep{TDApackage, otter2017roadmap, wadhwa2018tdastats}. Considering that the main purpose is statistical inference and online monitoring of the persistent diagrams of the scanned objects, it is more important to handle many large datasets in a short amount of time. \cite{zhao2022intrinsic} uses the \texttt{TDAstats} package, focusing mainly on the 0 and 1-dimensional features for this reason. In this paper, we re-examine these conclusions and packages with the most recent implementations.

A permutation test for whether an object has topological differences with respect to another was also developed by \cite{zhao2022intrinsic}, who presented limited simulation run length performance given the high computational requirements of PH calculations with algorithms available at the time. In this paper, we revisit this test and conduct a thorough run-length performance analysis with more recent methods of computation.

\section{Computation of Persistent Homology and Persistence Diagrams}
\label{Compute}

There are various existing software packages to calculate the persistent homology of a point cloud and its visual representation in {\bf \em persistence diagrams}, which will be our main focus for statistical inference. 

\begin{figure}[ht]
\centering
\begin{tabular}{ccc}
\includegraphics[width=0.285\textwidth]{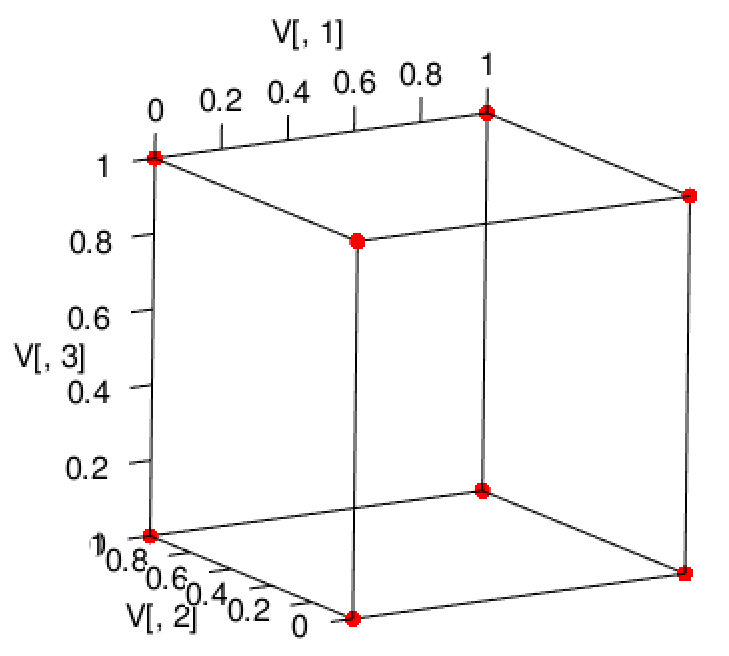}&\includegraphics[width=0.30\textwidth]{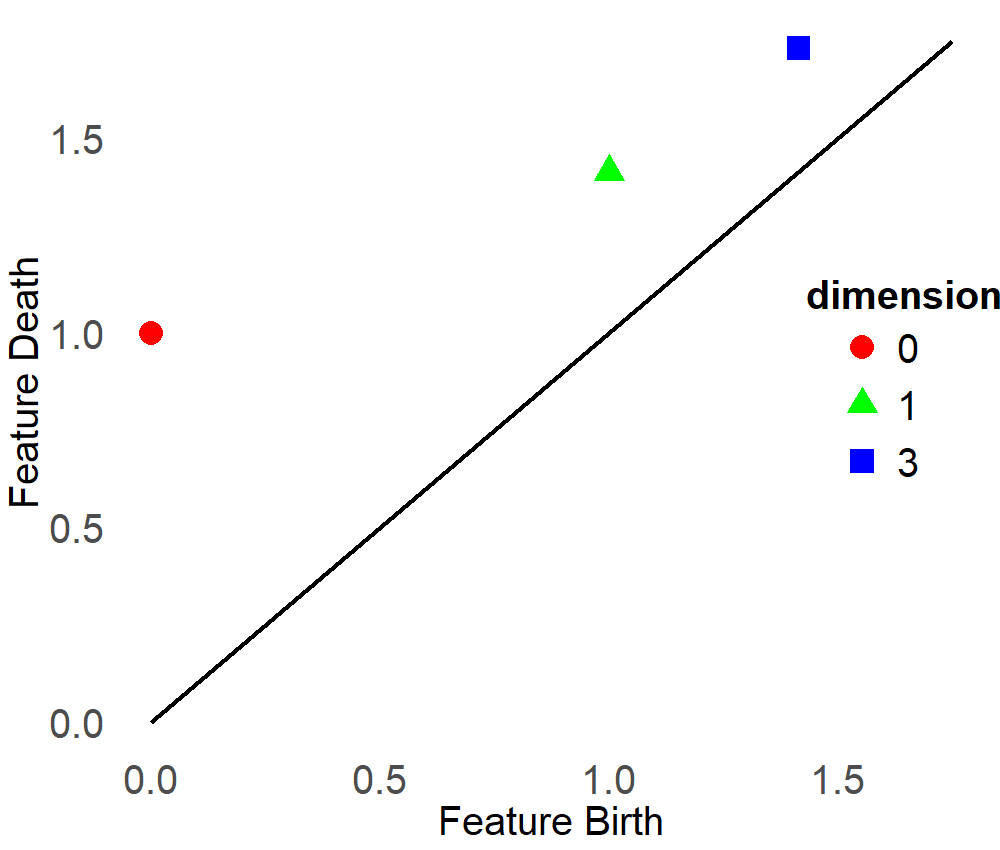}&\includegraphics[width=0.30\textwidth]{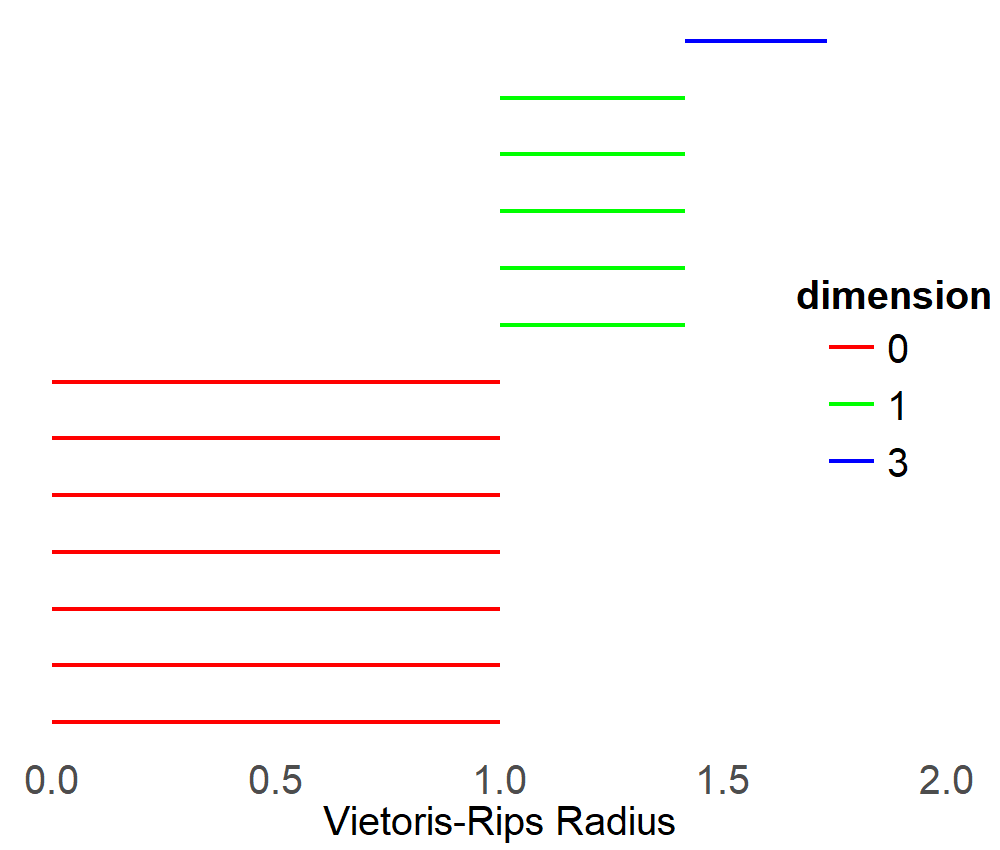}
\end{tabular}
\caption{Left: A point cloud consisting of the eight vertices of a cube (the red points). Middle: The persistence diagram of the point cloud in the left figure. Right: The barcode plot of the point cloud in the left figure.}
\label{fig:cube_example}
\end{figure}

Figure \ref{fig:cube_example} shows a simple example of a persistence diagram, where the point cloud consists only of the eight vertices of a cube (the red points in the left plot). The middle figure is the persistence diagram, where each point with coordinate $(x,y)$ represents a topological feature that is born at time $x$ and dies at time $y$. Although we can only see three points on this plot, there are more features that coincide with each other due to the perfect symmetry of the point cloud. This can be verified by the bar plot on the right, where each feature is represented by a bar starting at the time it is born and ending at the time it dies. The algorithm identifies seven 0-dimensional features, five 1-dimensional features, and one 3-dimensional feature. Features of the same dimension have exactly the same birth time and death time, thus overlapping with each other on the persistent homology diagram. The 0-dimensional features (vertices) are born at time 0 and die at time 1, when they are connected with their nearest neighbors. This is also exactly the time when the edges of the cube appear, and thus the 1-dimensional features (loops) are born. Further, at time $\sqrt{2}$, vertices that are diagonally opposite on the faces of the cube are connected, ``killing'' the loops surrounding those faces. Again, at the same time, the current simplicial complex looks like a sphere with a 2-dimensional void inside. However, if we examine the faces again, we will notice that now all four vertices on each face are pairwise connected, making a tetrahedron instead of triangles. That is why the feature that lives from time $\sqrt{2}$ to time $\sqrt{3}$ refers to the void inside the cube, identified as a 3-dimensional ``hole''.

We compare the performance of the two existing R software packages that compute persistent diagrams given a point cloud, namely packages {\tt TDA} and {\tt TDAstats} \citep{TDApackage, wadhwa2018tdastats}. 
The TDA package provides an R interface for the efficient algorithms of the C++ libraries
GUDHI, Dionysus, and PHAT \citep{TDApackage}. On the other hand, {\tt TDAstats} uses the Ripser C++ library to perform computations on a Vietoris-Rips complex \citep{wadhwa2018tdastats}.

Table \ref{PH_computing} lists the computing time of these two packages, with different numbers of points and varying maximum dimension of the topological features to be identified. The point cloud of 1,500 points represents a coarser skeletal approximation of the nominal ``egg'' part, as depicted in the leftmost panel of Figure \ref{TDAfig:Egg_Skeletons}, while the point cloud of 20,000 points is a remeshed version of the same part, shown in the leftmost panel of Figure \ref{fig:egg}, generated using the isotropic remeshing algorithm (IRA) described in \cite{an2025practical}.  As we can see from the results, the function {\tt alphaComplexDiag} in the  {\tt TDA} package computes the persistent homology much faster than the function {\tt calculate\_homology} from the {\tt TDAstats} package. Furthermore, the latter package cannot handle the computation of homologies of dimension higher than 0 for large point clouds. In various of our experiments, function {\tt calculate\_homology} crashed due to memory issues, probably due to the internal representation of the simplicial complex computation of the persistence diagram. In those cases where a crash occurs, an ``RSC'' was recorded to indicate R session crashing issues.

Given its computational efficiency and robustness, the {\tt TDA} package is adopted for subsequent analyses in this paper, whose goal is efficient statistical inference and online monitoring of persistence diagrams for complex geometries.

\begin{table}[ht]
\begin{center}
\begin{tabular}{c|c|c|c}
\hline
Number of & Maximum dimension & \multicolumn{2}{|c}{Computing time (seconds)} \\
\cline{3-4}
points & of features & {\tt TDA} & {\tt TDAstats} \\
\hline
1500 & 0 & 0.08116 & 0.29129 \\
1500 & 0,1 & 0.08116 & 4.61869 \\
1500 & 0,1,2 & 0.08116 & RSC \\
20000 & 0 & 2.48933 & 83.1171 \\
20000 & 0,1 & 2.48933 &  RSC \\
20000 & 0,1,2 & 2.48933 &  RSC \\
\hline
\end{tabular}
\caption{Computing time comparison between function {\tt alphaComplexDiag} in package {\tt TDA} and function {\tt calculate\_homology} in package {\tt TDAstats}. RSC stands for R session crashing issues due to exceeding memory during the computation. All experiments are tested on the same computer with 2.3 GHz 11th Gen Intel(R) Core(TM) i7 and 16G RAM.}
\label{PH_computing}
\end{center}
\end{table}%

\section{Distances Between Persistence Diagrams} \label{TDA.dist}

To compare the topological features of different point clouds and to eventually develop an SPC method on them, the first thing we need to do is to define a distance, or metric, between two persistence diagrams. There exist various proposed metrics in the space of persistence diagrams in the Statistics literature. \cite{robinson2017hypothesis} consider a distance metric that is analogous to the 2-Wasserstein distance between probability distributions (also known as the earth mover's distance). More specifically, they define
$$d_p(X,Y) = \left(\inf_{\phi:X\to Y}\sum_{x\in X}\|x-\phi(x)\|_p^p\right)^{1/p}$$
where $X$ and $Y$ are the two homology matrices, and $\phi$ is a bijection between features in $X$ and features in $Y$. The reason that $\phi$ can be assumed to be bijective is that the diagonal line in the persistence diagram can be seen as containing infinitely many points or features that die immediately after their birth. Therefore, should there be any point that does not have an image or pre-image under $\phi$, it can always be mapped to the closest point on the diagonal. When $p$ goes to infinity, the so-called bottleneck distance is obtained:
$$d_\infty(X,Y) = \inf_{\phi:X\to Y} \sup_{x\in X}\|x-\phi(x)\|_\infty$$
Both the p-Wasserstein distance and the bottleneck distance between two persistence diagrams can be computed with functions in the {\tt TDA} package. 

The {\tt TDAstats} package provides another distance between two persistence diagrams (function {\tt phom.dist}), which is claimed to be inspired by the Wasserstein distance. First, for each feature in $X$ and $Y$, its duration (death-birth) is calculated, resulting in two vectors of durations, denoted by $X_d$ and $Y_d$. Let $n_X$ and $n_Y$ be the length of $X_d$ and $Y_d$, respectively. Without loss of generality, suppose $n_X\leq n_Y$. Then, we add $n_Y-n_X$ number of zeros to the set $X_d$ so that $X_d$ and $Y_d$ are of equal length now. Finally, their distance is defined as 
\be
d(X,Y) = \sum_{i=1}^{n_Y} |x_{(i)}-y_{(i)}|
\label{PD_dist}
\ee
where $x_{(i)}$ and $y_{(i)}$ are the order statistics of (the augmented) $X_d$ and $Y_d$, respectively. 

When two persistence diagrams $X$ and $Y$ are very similar to each other, all distances introduced so far will be small. On the other hand, in an extreme case where $X$ contains a point, say $(x,y)=(0,3)$ and $Y$ contains a point such as $(x,y)=(6,9)$, the optimal bijection $\phi$ would match both of these points to the diagonal and $d_p(X,Y)$ would be significantly greater than zero. However, $d(X,Y)$ defined by {\tt phom.dist} in the {\tt TDAstats} package will give a distance of 0 since it only considers the duration of the features. This indicates that the {\tt phom.dist} function may tend to underestimate the distance between two persistence diagrams. It is worth noting that a single distance value is never meaningful, regardless of the metric used. Interpretations should only be made when comparable values, such as a null distribution, are available. Also, since all inter-part comparisons would be evaluated by the same distance, a consistent underestimation may not matter much. Due to these reasons, the apparent ``underestimation'' problem of the {\tt phom.dist} function is not as serious as it seems. 

Although the Wasserstein-like distances are more accurate and reliable, they need to solve an assignment problem via a linear program to find the optimal bijection $\phi$, which evidently requires more computation \citep{KerberMorozovNigmetov2017}. We use the {\tt phom.dist} function due to its computational advantage, and leave the task of finding a distance metric with a better balance between accuracy and computational complexity for further research. 

To compute the persistence diagrams themselves, we use the {\tt alphaComplexDiag} function in the {\tt TDA} package, which returns a matrix with three columns, where each row represents a detected topological feature. The first column denotes the dimension of each feature, while the next two columns record the birth time and death time accordingly. In this way, the homology matrix can be seen as a profile of the point cloud. (A noted limitation in {\tt TDA} is a potential bug in the {\tt alphaComplexDiag} function, which computes all $n-1$ lower-dimensional features regardless of the specified {\tt maxdimension} parameter. For instance, given a 3D point cloud, {\tt alphaComplexDiag} function computes 0-dimensional, 1-dimensional, and 2-dimensional features all at once. However, this is not a serious concern for our study, given its fast computation speed shown in Table \ref{PH_computing}. We can then selectively extract various dimensional features of our interests.) Upon extracting the output matrix from the {\tt alphaComplexDiag} function, we use the {\tt phom.dist} function from the {\tt TDAstats} to compute the distance between two persistence diagrams.


\section{A Permutation Test Based on the Persistence Diagrams of Parts} \label{TDA.permtest}

Having discussed distance metrics between persistence diagrams, we can statistically compare the topological structures between two sets of point clouds. In particular, we focus on the special case where one set contains multiple point clouds while the other set only has one point cloud. Let $\{X_1, X_2, ..., X_{m_0}\}$ be the first set of point clouds and $\{Y\}$ be the second set. We can think of the $X_i$'s and $Y$ being configuration matrices giving the coordinates of the points, and we do not assume they have the same number of points, or the same location/orientations. The null hypothesis we want to test is that $Y$ comes from the same distribution as the $X_i$'s, i=1, 2, ..., $m_0$. Using equation (\ref{PD_dist}), we obtain the distances between all pairs of  persistence diagrams, based on which the total pairwise distance within the first set can be calculated, namely
\be
d_\text{in-group} = \sum_{1\leq i<j\leq m_0} d(X_i,X_j)
\label{perm.stat}
\ee
This gives a measure of the similarity within group 1 and will be used as the test statistic. To obtain its null distribution, we randomly shuffle the labels of all point clouds. Since there is only one observation in group 2, we have $m_0+1$ permutations in total. Each point cloud will be assigned once to group 2, with all the others assigned to group 1. This gives $m_0+1$ values of the test statistic, which should come from the same distribution under the null hypothesis. These $m_0+1$ values form the null distribution, and the null hypothesis is rejected if the observed test statistic (computed based on the original allocation of the point clouds) is significantly {\em small} compared to this distribution. 

\begin{figure}[ht]
\centering
\begin{tabular}{ccc}
\includegraphics[width=0.22\textwidth]{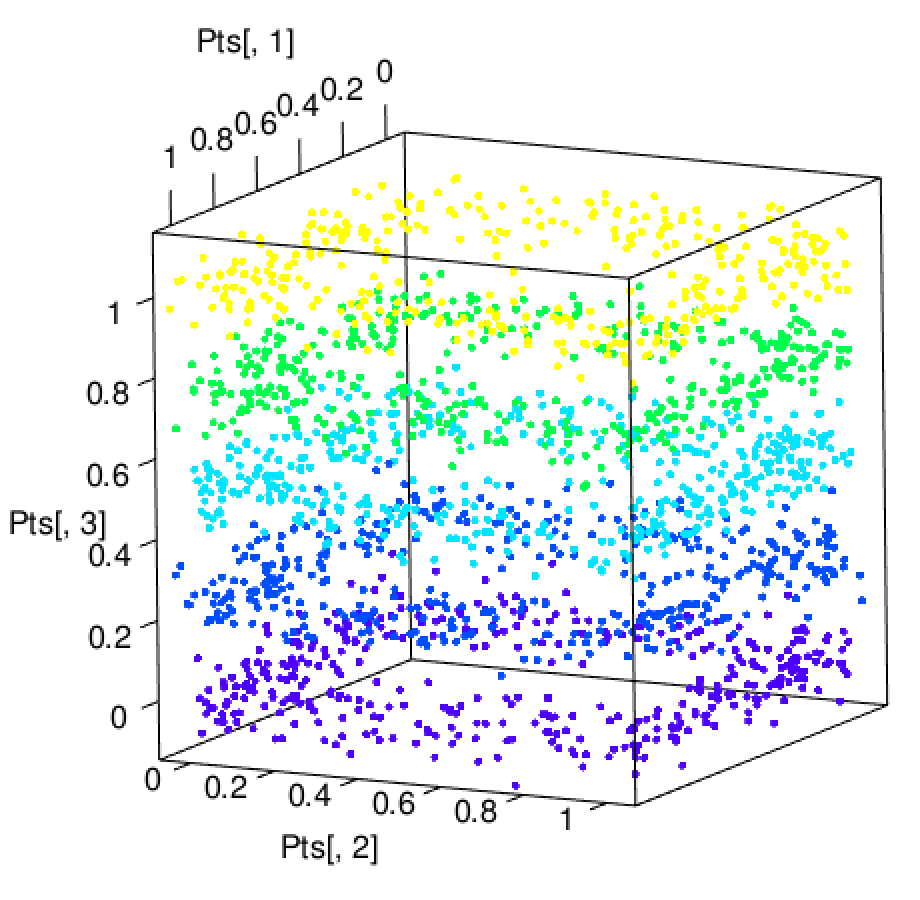}&\includegraphics[width=0.22\textwidth]{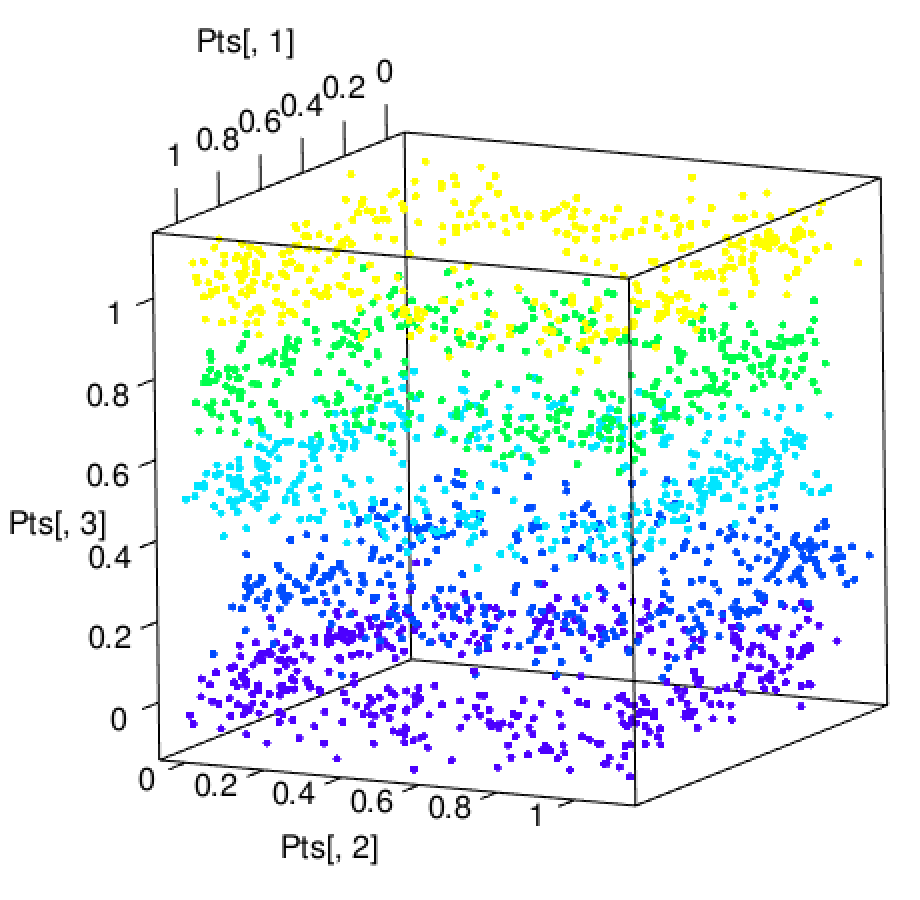}
&\includegraphics[width=0.5\textwidth]{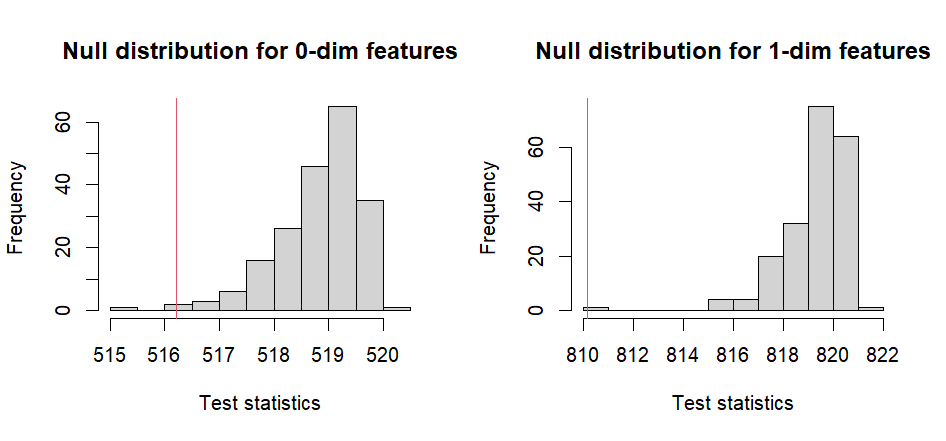}
\end{tabular}
\caption{From left to right: an in-control part where material is added layer by layer; an out-of-control part where the navy blue layer is slightly shifted along the y-axis by a magnitude of 0.1; the permutation test applied to the 0-dimensional features; the permutation test applied to the 1-dimensional features. The p-values for the two tests are 0.00498 and 0, respectively.}
\label{fig:5layers}
\end{figure}

To illustrate, Figure \ref{fig:5layers} shows a hypothetical scenario in Additive Manufacturing where a cube without the top and bottom faces (so it is homeomorphic to a cylinder) is constructed layer by layer. An example of the in-control part is displayed on the left, where the different layers are drawn with different colors. One potential defect in such a scenario is when the machine is not correctly aligned to the center, resulting in a slightly shifted layer, plotted in the second graph (the navy blue layer is shifted along the y-axis by a magnitude of 0.1). We simulate 200 in-control parts as group 1 (so $m_0$=200) and treat the out-of-control part as group 2. The permutation test described above gives the two null distributions displayed on the right, for 0- and 1-dimensional features, respectively. The observed test statistic is marked by the red vertical lines. The p-values are 0.00498 and 0, respectively, so we can conclude that the two groups differ.

\begin{figure}[ht]
\centering
\begin{tabular}{cc}
\includegraphics[width=0.4\textwidth]{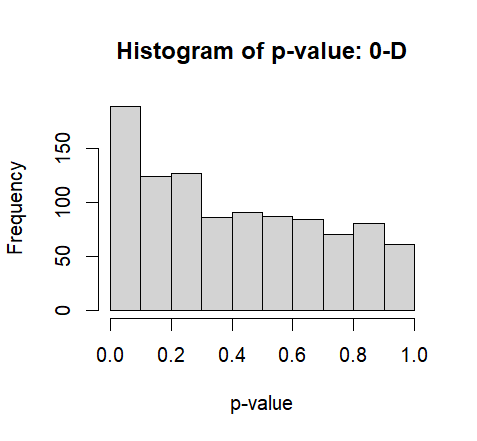}&\includegraphics[width=0.4\textwidth]{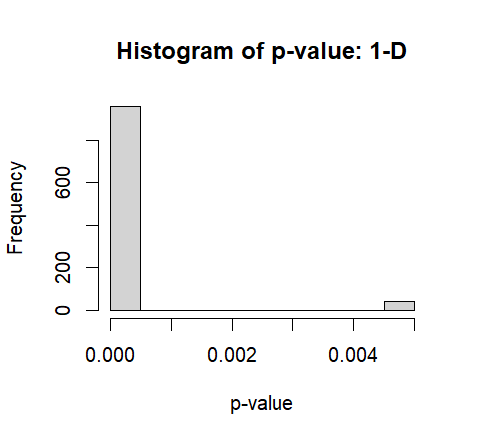} \\
\includegraphics[width=0.4\textwidth]{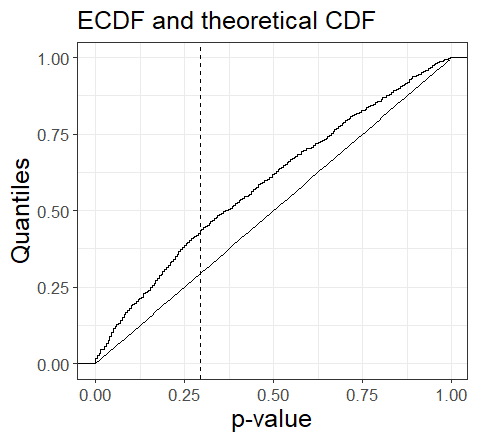}
&\includegraphics[width=0.4\textwidth]{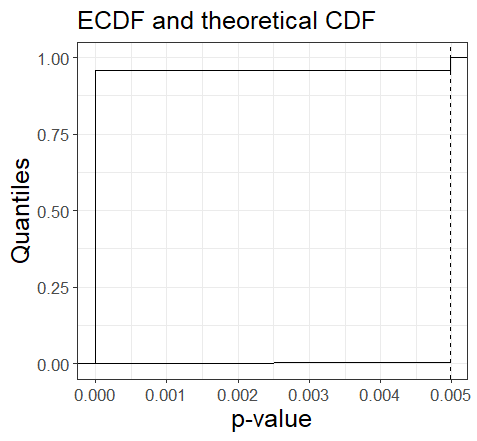}
\end{tabular}
\caption{Top row: histograms of the p-values from 1000 replications of the permutation test. Bottom row: ECDF plotted against the theoretical CDF, Uniform(0,1). Left: 0-dimensional features are being tested. Right: 0-dimensional features are being tested. The vertically dotted line in the plots on the bottom row represents the x-value where the maximum difference between the ECDF and the theoretical CDF occurs.}
\label{fig:5layers_permtests}
\end{figure}

To examine the general detection power of the permutation test, we repeat this process 1000 times. The top row in Figure \ref{fig:5layers_permtests} plots the histograms of the p-values from these 1000 tests. In the left plots, the 0-dimensional features are tested, and in the right plots, the 1-dimensional features are tested. Both histograms contain more smaller values than larger values of the p-values, which is expected since we are comparing an out-of-control part against a set of in-control parts, where the smaller p-values, the better. We also conducted the Kolmogorov-Smirnov (KS) test to compare this Empirical Cumulative Distribution Function (ECDF) and the theoretical Cumulative Distribution Function (CDF), a uniform distribution from 0 to 1 (see the bottom row in Figure \ref{fig:5layers_permtests}), which is what the p-value should follow when the hypothesis test cannot detect the difference. Both KS tests reject with high confidence the null hypothesis that the histograms plotted in the first row come from a Uniform(0,1) distribution, further confirming the detection power of the permutation test. In this example, this is particularly high for the 1-dimensional features.

We note that, although in this particular case, the 1-dimensional features seem to be more powerful at detecting different topological structures, this may not be the case for all types of defects. An 
alternative scheme is to combine the joint monitoring of  0-dimensional and 1-dimensional features and to use two charts in parallel, stopping a process when any one triggers an alarm and continuing otherwise. This has the danger of the well-known multiple comparisons problem, where an inflated number of alarms is introduced if the multiple tests are correlated. We investigate the performance of such a combined approach below.

\section{``Phase II'' SPC Scheme} \label{TDA.SPC}

We now naturally extend the permutation test introduced in the previous sections to a sequential test applied in the {\em online} ``Phase II'' of industrial SPC. Here we assume in-control parts from  ``Phase I'' are available, and serve as the first group of multiple point clouds for the test, $\{X_1, X_2, ..., X_{m_0}\}$. The topological features of the newly produced part will be tested against those of the parts in this group. In other words, whenever a new part $X_i$ is manufactured, $i>m_0$, the permutation test described in Section \ref{TDA.permtest} is applied to groups $\{X_1, X_2, ..., X_{m_0}\}$ and $\{X_i\}$. We continue this process until the null hypothesis is rejected. Note that this SPC scheme is a little different from the more common approaches, where the limits of the SPC charts are fixed and the test statistic changes as new parts are being produced. In our case, the test statistic remains constant regardless of part $X_i$ because it is only based on the reference set $\{X_1, X_2, ..., X_{m_0}\}$, so it does not change in phase II. On the other hand, the null distribution changes with $X_i$ since $m_0$ out of the total $m_0+1$ permutations will include $X_i$ in the first group. Hence, the limit, which is the $100\alpha$th percentile of the null distribution, changes from time to time.

\begin{figure}[ht]
\centering
\begin{tabular}{cc}
\includegraphics[width=0.47\textwidth]{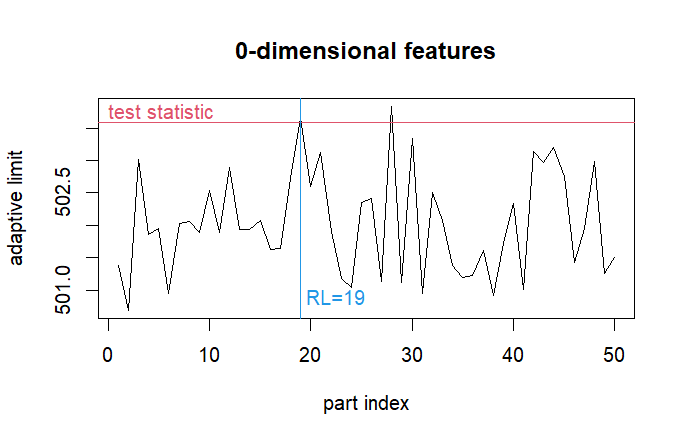}&\includegraphics[width=0.47\textwidth]{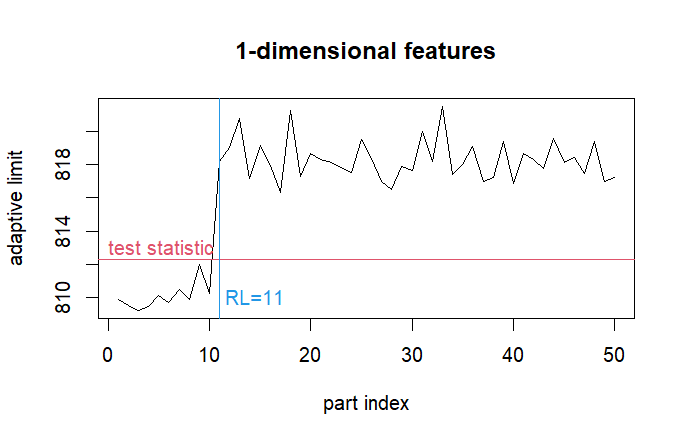}
\end{tabular}
\caption{SPC charts applied to the 0 and 1-dimensional features, respectively. $m_0=200$ ``Phase I'' parts (depicted in the middle plot, Figure \ref{TDAfig:parts}) are used. Defects (depicted in the right plot, Figure \ref{TDAfig:parts}) are introduced starting at part \#11. The adaptive limit is plotted in dashed black, while the fixed test statistic is in red. The blue vertical line denotes the run length. $\alpha=0.05$.}
\label{TDAfig:charts}
\end{figure}

\begin{figure}[ht]
\centering
\begin{tabular}{ccc}
\includegraphics[width=0.3\textwidth]{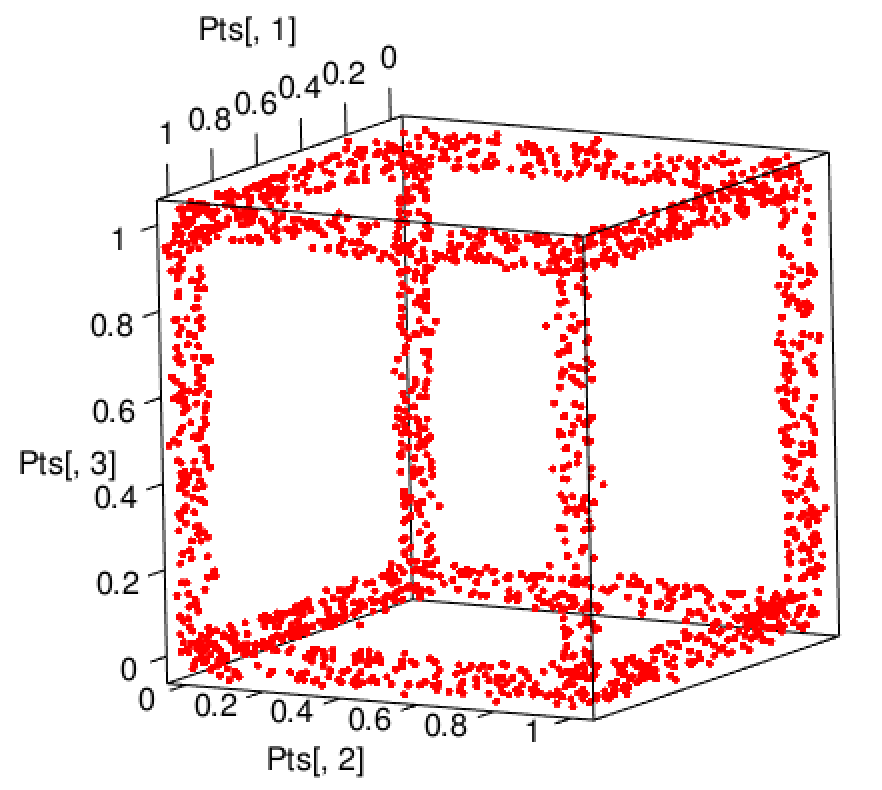}&\includegraphics[width=0.3\textwidth]{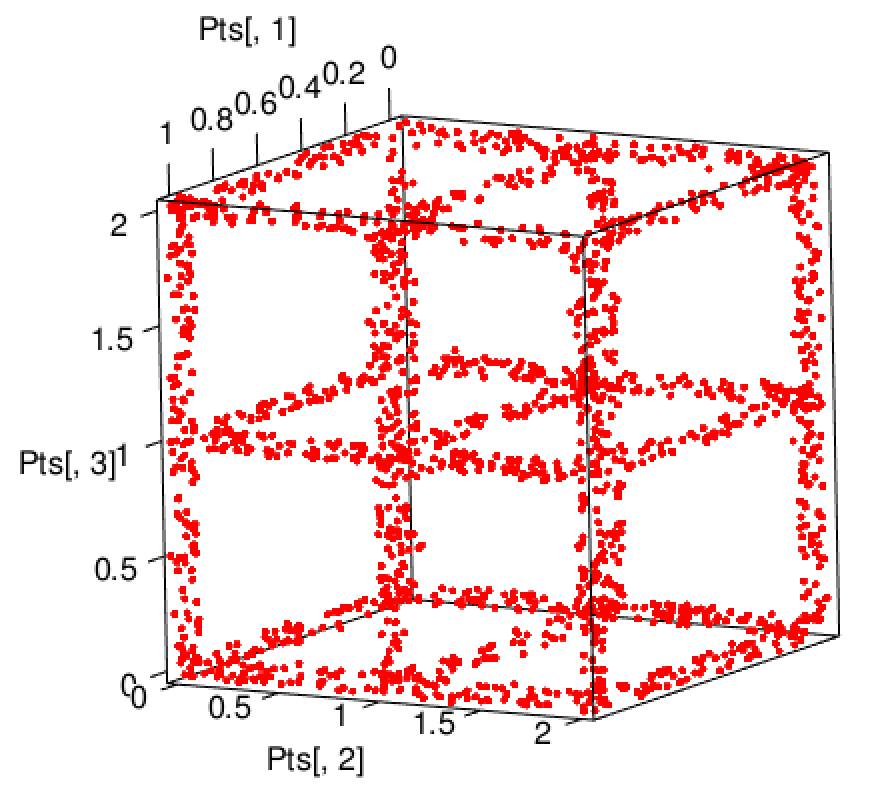} 
& \includegraphics[width=0.3\textwidth]{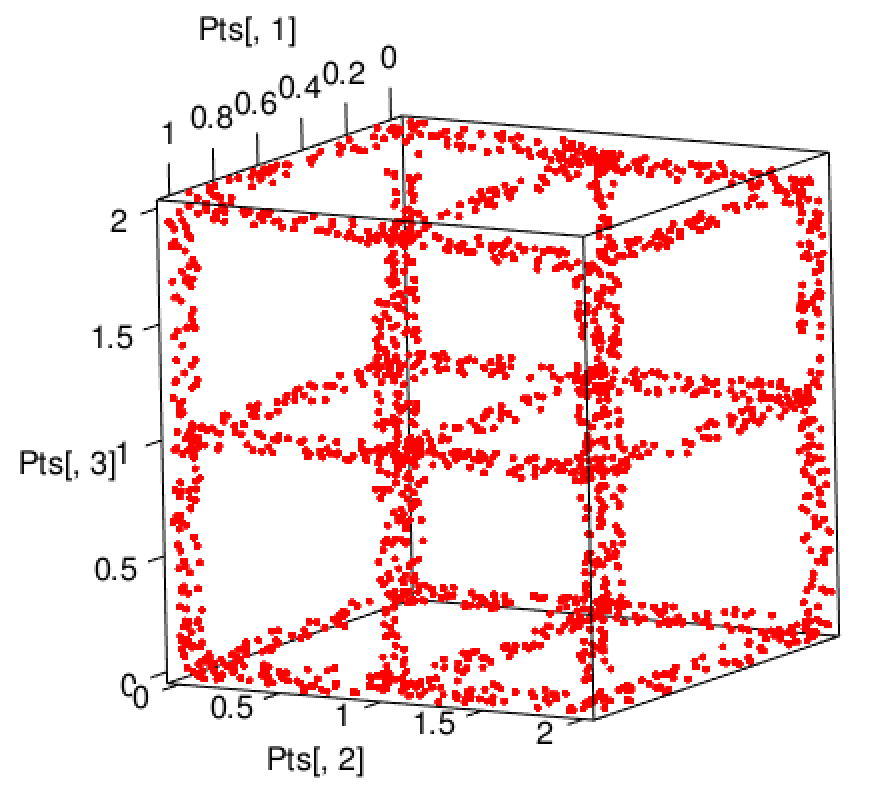}
\end{tabular}
\caption{Left: a single cube as an example of an in-control part. Middle: an in-control 4-cube lattice. Right: an out-of-control 4-cube lattice with a collapsed top edge. The simulated noise in the 3D coordinates is given by a $N\sim(0,\sigma^2)$, i.e., the noise is isotropic Gaussian.}
\label{TDAfig:parts}
\end{figure}

Figure \ref{TDAfig:charts} shows an example of a proposed SPC chart based on the sequential application of the previously presented permutation test. The two figures are results from using the 0-dimensional and 1-dimensional features, respectively. Two hundred Phase I (in-control) parts are used, with a simple lattice structure where four cubes are stacked together (middle plot in Figure \ref{TDAfig:parts}). The type of assignable cause being tested is a potential type of defect when the material melts and fails to hold the desired shape, resulting in a slightly collapsed top edge, as the right plot in Figure \ref{TDAfig:parts} shows. In this exercise, the defective parts are introduced starting at part \#11, and the 1-dimensional features detect it immediately, whereas the left chart with the 0-dimensional features is not as sensitive as the detection with the 1-dimensional features. 

By repeatedly applying the permutation test at each step, we have a probability $\alpha$ of falsely declaring an in-control part to be out-of-control. Then the theoretical in-control run length, the number of parts until the first alarm when all parts are in-control, follows a geometric distribution with probability $\alpha$ if we assume the sequential tests are independent of each other. Considering the large portion of null distribution observations affected by the new part, this is reasonable if the newly produced parts can be assumed to be independent of each other.

We conduct a run length simulation analysis for three typical types of parts to examine the in-control run length (RL) performance of the proposed SPC chart. Tables \ref{TDA_RL_IC_dim0} and \ref{TDA_RL_IC_dim1} summarize the run length performance from 10000 replications, using 0-dimensional and 1-dimensional features, respectively. One pattern we can identify from the tables is that when $m_0$ is not very large ($m_0 \leq 150$), the observed average run length (ARL) and standard deviation of the run length (SDRL) tend to be larger than their theoretical values, which are 20 and 19.49, respectively, when $\alpha=0.05$. This is due to the fact that the null distribution has almost the same size as $m_0$, thus, it is not large enough to guarantee the nominal type I error. As $m_0$ increases from 100 to 200, all in-control cases have run length behavior close to the nominal geometric distribution with probability $\alpha=0.05$. For this reason, we recommend using a reference or ``phase I'' set of observations of at least $m_0=200$ parts for SPC purposes. 

\begin{table}[ht]
\begin{center}
\begin{tabular}{c|c|c|c|c|c}
\hline
 Part & $m_0=100$ & $m_0=150$ & $m_0=200$ & $m_0=250$ & $m_0=300$ \\
 \hline
 Single cube & 25.28(31.41) & 21.51(24.13) & 22.12(24.12) & 20.57(21.67) & 21.05(21.67) \\
 5-layered cube & 25.38(35.77) & 21.23(23.60) & 22.45(24.69) & 20.95(22.52) & 21.43(22.55) \\
 4-cubed lattice & 25.01(31.66) & 21.24(23.60) & 22.40(24.45) & 20.53(21.47) & 21.37(22.56) \\
\hline
\end{tabular}
\end{center}
\caption{In-control average (and standard deviation) run length performance for a chart monitoring only the 0-dimensional features, cube parts. Results are obtained from 10000 replications, $\alpha=0.05$.}
\label{TDA_RL_IC_dim0}
\end{table}%

\begin{table}[ht]
\begin{center}
\begin{tabular}{c|c|c|c|c|c}
\hline
 Part & $m_0=100$ & $m_0=150$ & $m_0=200$ & $m_0=250$ & $m_0=300$ \\
 \hline
 Single cube & 25.45(32.16) & 21.28(23.99) & 22.37(25.40) & 20.91(22.50) & 21.31(22.15) \\
 5-layered cube & 24.68(30.66) & 21.42(24.19) & 22.13(23.75) & 20.78(21.90) & 21.74(22.80) \\
 4-cubed lattice & 25.20(31.73) & 21.28(23.36) & 22.44(24.67) & 20.72(22.39) & 21.33(22.46) \\
\hline
\end{tabular}
\end{center}
\caption{In-control average (and standard deviation) run length performance for a chart monitoring only the 1-dimensional features, cube parts. Results are obtained from 10000 replications, $\alpha=0.05$.}
\label{TDA_RL_IC_dim1}
\end{table}%

We also investigated the multiple testing problem by jointly monitoring 0-dimensional and 1-dimensional features on the same set of objects, where an alarm is triggered if either the 0-dimensional or 1-dimensional control chart signals, and monitoring continues as long as both charts do not trigger alarms. Whenever this is done, there is an increased risk of falsely rejecting the null hypothesis of an in-control process. Evidently, when multiple testing, the ARL and SDRL are expected to be lower than their respective single-feature benchmark values of 20 and 19.49. As reported in Table \ref{TDA_RL_IC_dim0_and_dim1}, the run length results for this joint testing scenario reveal that both ARL and SDRL values are approximately halved compared to those obtained from separate testing of 0-dimensional and 1-dimensional features, as shown in Tables \ref{TDA_RL_IC_dim0} and \ref{TDA_RL_IC_dim1}, highlighting the impact of multiple testing on control chart performance. This suggests that the two SPC charts based on the 0-dimensional and 1-dimensional features are independent, since their combined ARL, assuming the user sets the two charts to have the same in control ARLs (namely, $ARL_0^1=ARL_0^2$) equals $ARL_0^C = \frac{ARL_0^1}{2}=\frac{ARL_0^2}{2}$.

\begin{table}[ht]
\begin{center}
\begin{tabular}{c|c|c|c|c|c}
\hline
 Part & $m_0=100$ & $m_0=150$ & $m_0=200$ & $m_0=250$ & $m_0=300$ \\
 \hline
 Single cube & 12.07(13.24) & 10.79(10.89) & 11.34(11.49) & 10.62(10.46) & 11.01(10.97) \\
 5-layered cube & 12.14(13.19) & 11.00(11.31) & 11.67(11.86) & 11.09(11.16) & 11.45(11.39) \\
 4-cubed lattice & 11.87(13.33) & 10.61(10.81) & 11.17(11.38) & 10.36(10.27) & 10.73(10.44) \\
\hline
\end{tabular}
\end{center}
\caption{In-control average (and standard deviation) run length performance of a combined approach where 0-dimensional and 1-dimensional features are jointly monitored, cube parts. Results are obtained from 10000 replications, $\alpha=0.05$ for each feature.}
\label{TDA_RL_IC_dim0_and_dim1}
\end{table}%

\begin{figure}[ht]
\centering
\begin{tabular}{ccc}
\includegraphics[width=0.3\textwidth]{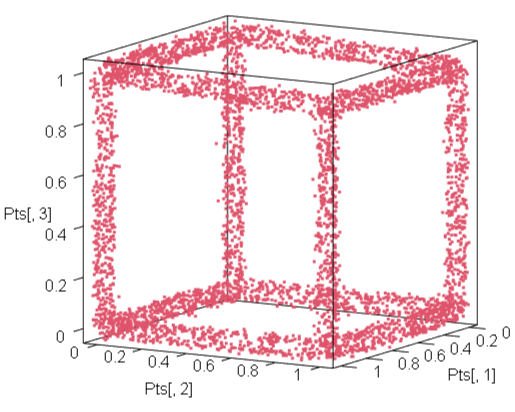}&\includegraphics[width=0.3\textwidth]{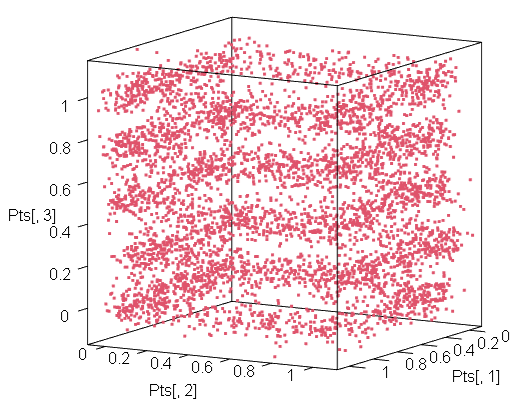} 
& \includegraphics[width=0.3\textwidth]{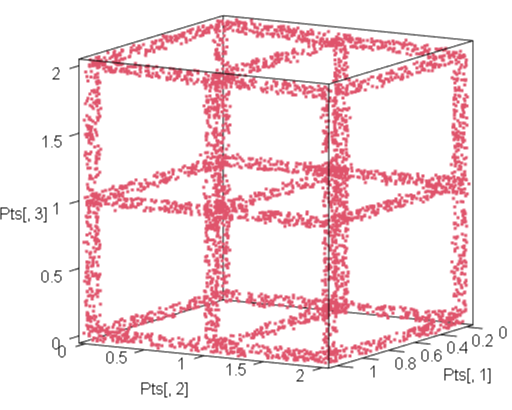}
\end{tabular}
\caption{Left: an out-of-control single cube with a collapsed top edge. Middle: an out-of-control 5-layered cube with a shifted layer. Right: an out-of-control 4-cube lattice with a collapsed top edge. The severity of the defect is $\delta = 0.05$, corresponding to the drop in height in a corner in the first and last part, and the shift of a layer in the middle part. The simulated noise in the 3D coordinates is given by a $N\sim(0,\sigma^2)$, i.e., the noise is isotropic Gaussian.}
\label{TDAfig:OOCparts}
\end{figure}

Figure \ref{TDAfig:OOCparts} illustrates examples of defects introduced in three parts, each with a severity of $\delta = 0.05$, corresponding to a deviation of 0.05 unit from the nominal geometry. Table \ref{TDA_AL_IC_OC} presents the out-of-control run length results, derived from 10,000 replications for each part, demonstrating the exceptional performance of the proposed SPC chart. The chart detects defects almost immediately when monitoring 1-dimensional features, highlighting their high sensitivity. In contrast, detection using 0-dimensional features is significantly less effective, resulting in the combined approach’s average run length (ARL) and standard deviation of run length (SDRL) being primarily driven by the detection performance of the 1-dimensional features.

\begin{table}[ht]
\begin{center}
\begin{tabular}{c|c|c|c}
\hline
 Defects ($\delta = 0.05$) & 0-dimensional & 1-dimensional & Combined \\
 \hline
 Single cube & 15.485(15.833) & 1.002(0.039) & 1.002(0.039)  \\
 5-layered cube &  15.976(16.671) & 1.247(0.574) & 1.240(0.563) \\
 4-cubed lattice & 14.698(15.278) & 1.029(0.175) & 1.029(0.174) \\
 \hline
\end{tabular}
\end{center}
\caption{Out-of-control average (and standard deviation)  run length performance of the proposed SPC scheme, estimated from 10000 replications. The performance of both 0-dimensional and 1-dimensional charts is individually and jointly assessed. $m_0=200, \alpha=0.05$.}
\label{TDA_AL_IC_OC}
\end{table}

\section{Simulation Case Study}
\label{TDA.CaseStudy}

\begin{figure}[ht]
\centering
\begin{tabular}{ccc}
\includegraphics[width=0.525\textwidth]{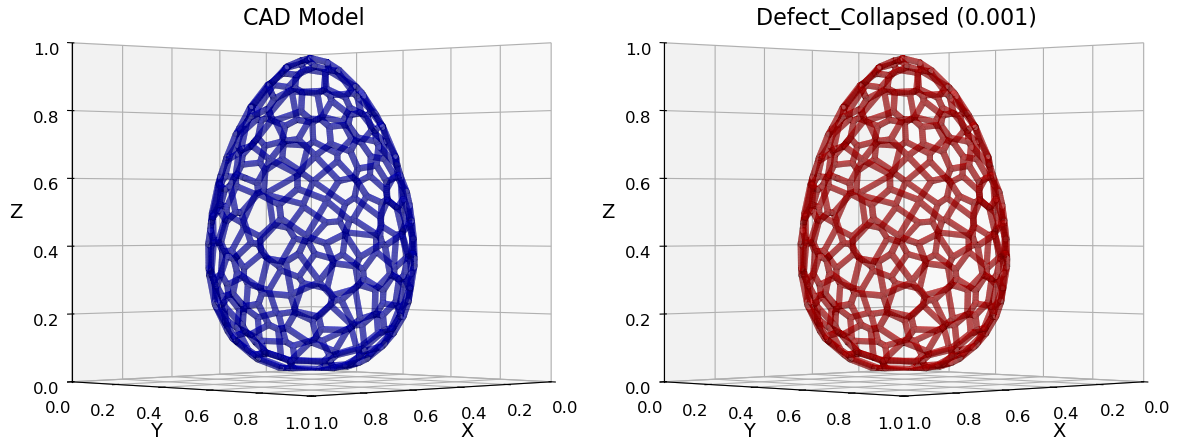}&\includegraphics[width=0.2\textwidth]{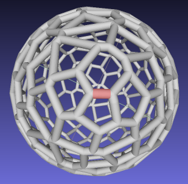}
&\includegraphics[width=0.2\textwidth]{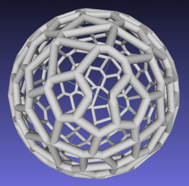}
\end{tabular}
\caption{From left to right: an in-control ``egg'' part; an out-of-control egg part with a height to width defect (``collapse'' defect), where the z-axis is reduced by a magnitude of 0.001 and the x-axis and y-axis are pushed outward by a magnitude of 0.001; the top view of an in-control egg part with the red section indicating the location of the missing strut defect that is introduced in the part at the rightmost panel; an out-of-control egg part with a missing strut defect. }
\label{fig:egg}
\end{figure}

To more systematically evaluate the effectiveness of the proposed topological statistical process control (SPC) method, we conduct a case study using synthetic data derived from the computer-aided design (CAD) model shown in the leftmost panel of Figure \ref{fig:egg} \citep{scimone2022statistical}. To assess the method’s sensitivity to structural anomalies in this type of part (which we will refer as ``egg''), we rescale the CAD model into a unity cube and simulate two common defect types: (i) a ``collapsed'' structure (where the height is reduced and the width increased by an equal amount), and (ii) a missing strut, as shown in the second and fourth panels in Figure \ref{fig:egg}.

\begin{figure}[!ht]
\centering
\begin{tabular}{ccc}
\includegraphics[width=0.3\textwidth]{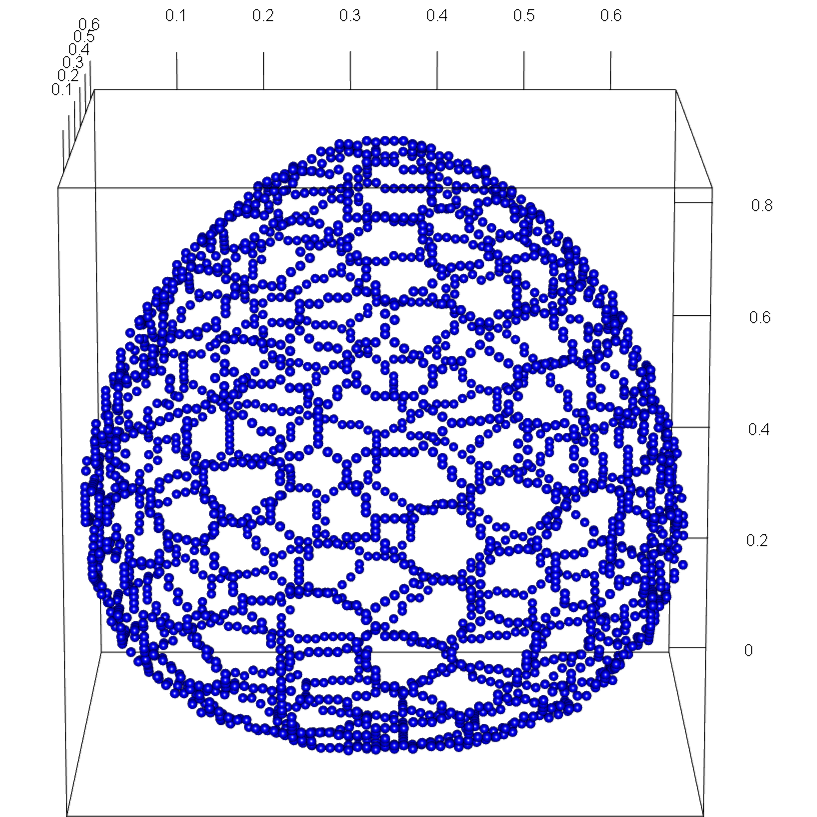}&\includegraphics[width=0.3\textwidth]{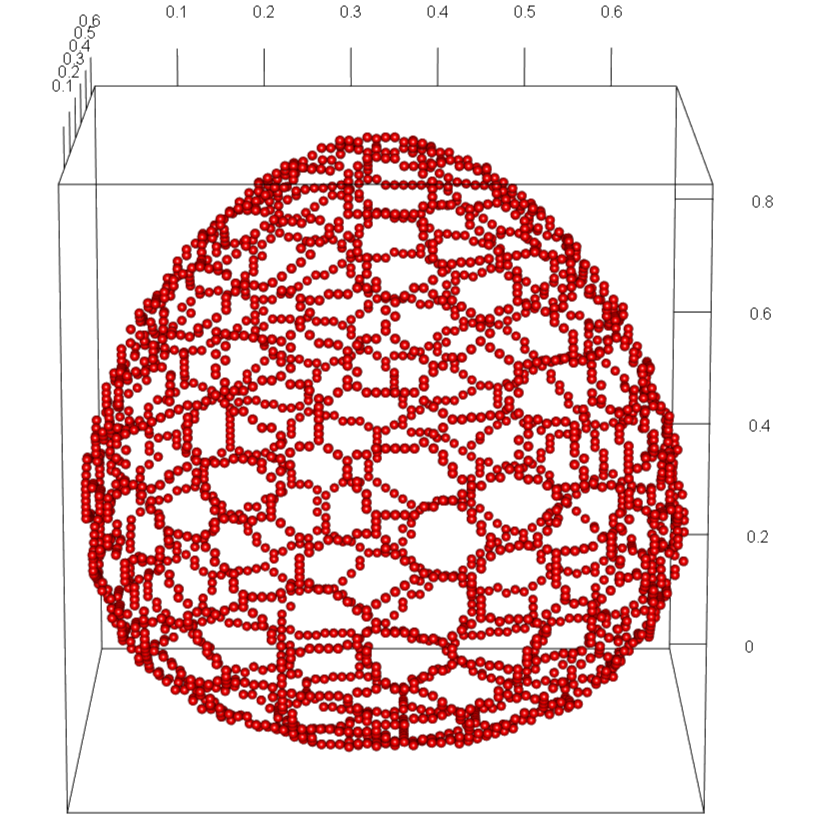} 
& \includegraphics[width=0.3\textwidth]{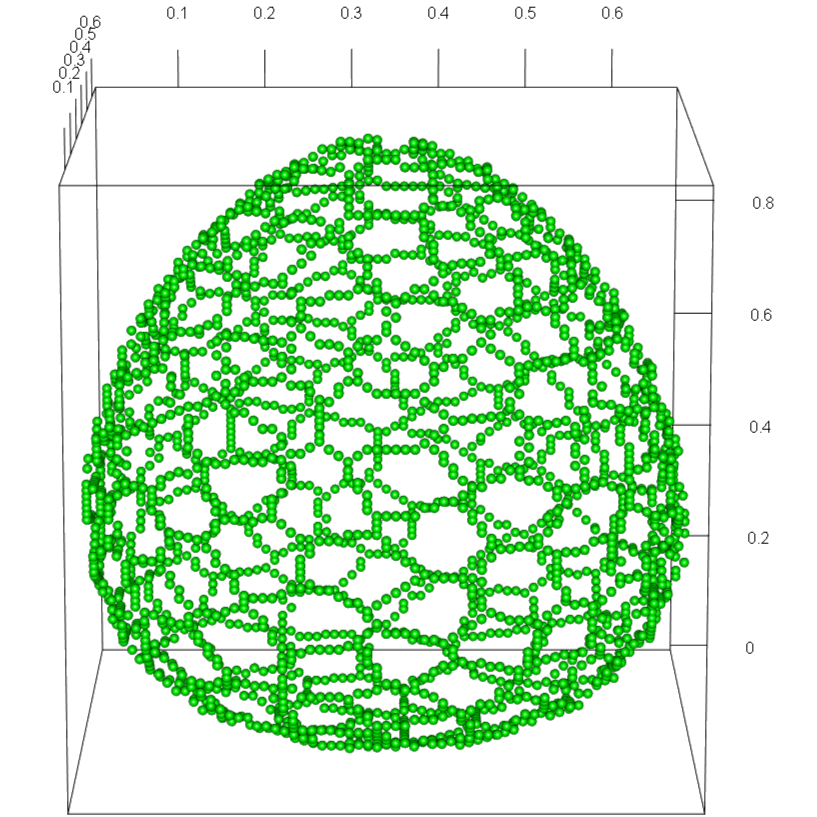}
\end{tabular}
\caption{Skeletons of the nominal and defective ``egg'' parts. Left: the skeleton of an in-control egg. Middle: the skeleton of an out-of-control egg with a collapse-type defect (decreased height, increased width). Right: the skeleton of an out-of-control egg with a missing strut type of defect. Then the simulated noise in the 3D coordinates is added using a $N\sim(0,\sigma^2)$, i.e., the noise is isotropic Gaussian.}
\label{TDAfig:Egg_Skeletons}
\end{figure}

Due to the high complexity of the geometry, the original CAD model has 357,982 vertices. Therefore, a dimensionality reduction technique known as skeletonization, developed by \cite{lee1994building}, is employed using the Python package {\tt skimage} to process the lattice-structured geometry \citep{van2014scikit}. This method efficiently compresses the original point cloud, reducing its size from approximately 357,000 to less than 3000 points, while retaining the essential topological characteristics of the object, particularly suitable for the two types of defects under our consideration. The resulting skeletal representations are illustrated in Figure \ref{TDAfig:Egg_Skeletons}.

To evaluate the in-control performance of the proposed permutation-based SPC chart, a run length analysis was conducted on sequences of ``egg'' parts under varying noise levels, focusing on 0-dimensional and 1-dimensional features, both independently and jointly, with results from 10,000 replications summarized in Tables \ref{TDA_RL_EggIC_dim0} through \ref{TDA_RL_EggIC_dim0_and_dim1}. Similar results as before, when $m_0$ is not very large ($m_0 \leq 150$), the observed average RL (ARL) and standard deviation of the RL (SDRL) tend to be larger than their theoretical values, which are 20 and 19.49, respectively. However, as $m_0$ increases from 100 to 200, the in-control run length behavior aligns closely with the nominal geometric distribution at a significance level of $\alpha = 0.05$. Consequently, $m_0 = 200$ was selected as the parameter for subsequent out-of-control simulations in the TDA-based SPC scheme, ensuring robust and reliable performance for defect detection in complex geometries.

\begin{table}[ht]
\begin{center}
\begin{tabular}{c|c|c|c|c|c}
\hline
 noise level ($\sigma$) & $m_0=100$ & $m_0=150$ & $m_0=200$ & $m_0=250$ & $m_0=300$ \\
 \hline
 0.001 & 25.17(32.94) & 21.78(25.13) & 22.50(24.78) & 20.58(21.98) & 21.61(22.70) \\
 0.005 & 25.37(32.91) & 21.54(23.93) & 22.52(24.64) & 20.81(22.01) & 21.20(22.44) \\
 0.01 & 24.72(31.19) & 21.19(24.37) & 21.80(23.29) & 21.14(22.61) & 21.39(22.41) \\
 0.1 & 24.84(29.72) & 21.68(25.08) & 22.38(24.53) & 20.84(22.20) & 21.54(22.87) \\
\hline
\end{tabular}
\end{center}
\caption{In-control average (and standard deviation) run length performance for the ``egg'' parts, when only 0-dimensional features are monitored. Results are obtained from 10000 replications, $\alpha=0.05$.}
\label{TDA_RL_EggIC_dim0}
\end{table}%

\begin{table}[ht]
\begin{center}
\begin{tabular}{c|c|c|c|c|c}
\hline
 noise level ($\sigma$) & $m_0=100$ & $m_0=150$ & $m_0=200$ & $m_0=250$ & $m_0=300$ \\
 \hline
 0.001 & 24.84(30.52) & 21.49(23.90) & 22.26(23.69) & 20.74(22.21) & 21.74(22.54) \\
 0.005 & 24.84(31.70) & 21.14(23.28) & 22.68(24.49) & 21.08(22.01) & 21.37(22.21) \\
 0.01 & 24.78(29.98) & 20.96(23.80) & 22.34(24.58) & 21.02(22.58) & 21.40(22.63) \\
 0.1 & 25.20(31.78) & 21.40(23.87) & 22.72(25.24) & 20.62(21.73) & 21.56(22.90) \\
\hline
\end{tabular}
\end{center}
\caption{In-control average (and standard deviation) run length performance for the ``egg'' parts, when only 1-dimensional features are monitored. Results are obtained from 10000 replications, $\alpha=0.05$.}
\label{TDA_RL_EggIC_dim1}
\end{table}%

\begin{table}[ht]
\begin{center}
\begin{tabular}{c|c|c|c|c|c}
\hline
 noise level & $m_0=100$ & $m_0=150$ & $m_0=200$ & $m_0=250$ & $m_0=300$ \\
 \hline
 0.001 & 11.45(12.58) & 10.47(10.78) & 10.96(11.14) & 10.16(9.92) & 10.89(10.77) \\
 0.005 & 12.22(13.47) & 10.96(11.30) & 11.51(11.80) & 10.77(10.51) & 11.01(10.84) \\
 0.01 & 12.76(14.26) & 11.43(11.96) & 12.01(12.43) & 11.65(11.77) & 11.81(11.85) \\
 0.1 & 13.38(14.66) & 11.96(12.74) & 12.95(13.03) & 11.83(11.97) & 12.25(12.36) \\
\hline
\end{tabular}
\end{center}
\caption{In-control average (and standard deviation) run length performance for the ``egg'' parts, when both 0-dimensional and 1-dimensional features are jointly monitored. Results are obtained from 10000 replications, $\alpha=0.05$ for each feature.}
\label{TDA_RL_EggIC_dim0_and_dim1}
\end{table}%

Lastly, Tables \ref{TDA_AL_OC_C} and  \ref{TDA_AL_OC_MS} report the out-of-control run length results with 10000 replications for the collapse and missing strut defects, respectively. The severity of the collapse defect was parametrized as multiples $k$ of the standard deviation of the noise ($k \sigma$). In contrast, there is no possible parametrization for the missing strut defect, being a binary variable, thus, its performance was assessed simply under varying noise levels (taking into consideration the size of the point cloud). The results indicate that, as expected, the ARL and SDRL values decrease with larger changes in the parts (i.e., as $k$ increases), with 1-dimensional features enabling faster defect detection compared to 0-dimensional features in these particular types of defects of this particular part. Under moderate noise conditions, the proposed TDA-based SPC method demonstrates rapid detection of both defect types, highlighting its robustness and efficacy for monitoring complex geometries in industrial applications.

\begin{table}[ht]
\begin{center}
\begin{tabular}{c|c|c|c}
\hline
 change ($k \sigma$) & 0-dimensional & 1-dimensional & Combined \\
 \hline
 2 $\sigma$ & 1(0) & 1(0) & 1(0)  \\
 1 $\sigma$ &  2.006(1.448) & 1.022(0.153) & 1.012(0.111) \\
 1/2 $\sigma$ & 6.576(6.480) & 3.239(2.869) & 2.384(1.874) \\
 1/3 $\sigma$ & 11.821(12.253) & 9.559(9.703) & 5.434(5.120) \\
 1/5 $\sigma$ & 18.097(19.403) & 16.545(17.504) & 8.668(8.724) \\
 \hline
\end{tabular}
\end{center}
\caption{Out-of-control average (and standard deviation) run length performance of the ``egg'' parts for the ``collapse'' type defect, estimated from 10000 replications. Both 0-dimensional and 1-dimensional features are individually and jointly monitored, $m_0=200, \alpha=0.05$ for each feature.}
\label{TDA_AL_OC_C}
\end{table}

\begin{table}[ht]
\begin{center}
\begin{tabular}{c|c|c|c}
\hline
  noise level ($\sigma$)  & 0-dimensional & 1-dimensional & Combined \\
 \hline
 0.0005 & 9.144(9.775) & 1.053(0.237) & 1.048(0.226)  \\
 0.001 & 16.46(17.731) & 2.377(1.884) & 2.156(1.643) \\
 0.002 & 19.285(21.456) & 8.222(8.321) & 5.826(5.556) \\
 0.003 & 18.880(20.377) & 13.841(14.530) & 8.041(7.868) \\
 0.005 & 22.436(23.809) & 15.356(15.737) & 9.385(9.195) \\
 \hline
\end{tabular}
\end{center}
\caption{Out-of-control average (and standard deviation) run length performance of the ``egg'' parts for the missing strut defect, estimated from 10000 replications. Both 0-dimensional and 1-dimensional features are individually and jointly monitored, $m_0=200, \alpha=0.05$ for each feature.}
\label{TDA_AL_OC_MS}
\end{table}

\section{Comparing TDA SPC With Spectral SPC}
\label{TDAvsLB}

To assess the efficacy of the proposed TDA-based SPC method for monitoring lattice-structured objects, this section compares its performance against the spectral SPC method presented in \cite{an2025practical} and \cite{zhaoEDC_Tech}, using the complex ``egg'' part from the prior simulation study. To ensure comparability, the TDA-based SPC method omits skeletonization, as the spectral SPC method requires mesh data for finite element method (FEM)-estimated Laplace-Beltrami (LB) spectra. The isotropic remeshing algorithm (IRA) was applied to the noise-free meshes of the nominal and defective ``egg'' parts (collapse and missing strut defects), as shown in Figure \ref{fig:egg}, producing remeshed meshes with approximately 3,000 vertices, to which various noise levels were subsequently added. 

Tables \ref{TDAvsSpectral_AL_OC_C} and \ref{TDAvsSpectral_AL_OC_MS} present the out-of-control average run length (ARL) and standard deviation of run length (SDRL) results from 5,000 replications for the collapse and missing strut defects in the ``egg'' part, respectively, with defect severity parameterized as described in the previous section, highlighting the superior performance of the proposed TDA-based SPC method in detecting defects in complex AM geometries. For large defect severities (top 2 rows of both tables), the TDA-based SPC method achieves instantaneous detection across 0-dimensional, 1-dimensional, and combined features, outperforming the spectral method’s higher ARL and SDRL values. For moderate severities (3rd and 4th rows of both tables), the TDA-based method using 1-dimensional features and combined features maintains lower ARL values compared to the spectral method, with 1-dimensional features consistently outperforming 0-dimensional features. For small severities (last row of both tables), however, the spectral method shows superior run-length performance compared to the TDA-based method in the ``collapse'' type of defects, while it remains slightly worse performance compared to the TDA-based method in the missing strut type of defects. \vspace{1cm}

\begin{table}[ht]
\begin{center}
\begin{tabular}{c|c|c|c|c}
\hline
 & \multicolumn{3}{|c|}{TDA} & Spectral \\
 \hline
 change ($k \sigma$) & 0-dimensional & 1-dimensional & Combined & 1 $\sim$ 15 LBS\\
 \hline
 2 $\sigma$ & 1(0) & 1(0) & 1(0) & 1.450(0.498) \\
 1 $\sigma$ &  1(0) & 1(0) & 1(0) & 1.461(0.499) \\
 1/2 $\sigma$ & 1.262(0.576) & 1.002(0.045) & 1.001(0.024) & 1.505(0.500) \\
 1/3 $\sigma$ & 3.976(3.775) & 1.353(0.689) & 1.245(0.550) & 1.876(0.514) \\
1/5 $\sigma$ & 8.196(8.225) & 5.640(5.221) & 3.973(3.600) & 3.414(1.318) \\
 \hline
\end{tabular}
\end{center}
\caption{Out-of-control average (and standard deviation) run length performance of the ``egg'' parts for the ``collapse'' type defect without skeletonization, estimated from 5000 replications. For the TDA-based SPC method, both 0-dimensional and 1-dimensional features are individually and jointly monitored, and for the spectral SPC method, the first 15 LB eigenvalues are used. $m_0=200, \alpha=0.05$ for both methods.}
\label{TDAvsSpectral_AL_OC_C}
\end{table}%

\begin{table}[ht]
\begin{center}
\begin{tabular}{c|c|c|c|c}
\hline
 & \multicolumn{3}{|c|}{TDA} & Spectral \\
 \hline
  noise level ($\sigma$)  & 0-dimensional & 1-dimensional & Combined & 1 $\sim$ 15 LBS \\
 \hline
 0.0005 & 1(0) & 1(0) & 1(0) & 1.459(0.499) \\
 0.001 & 1(0) & 1(0) & 1(0) & 1.480(0.500) \\
 0.002 & 1.008(0.089) & 1(0) & 1(0) & 1.714(0.462) \\
 0.003 & 1.445(0.845) & 1.093(0.320) & 1.042(0.212) & 1.992(0.556) \\
 0.005 & 2.516(2.026) & 2.584(2.118) & 1.783(1.225) & 2.957(1.046) \\
\hline
\end{tabular}
\end{center}
\caption{Out-of-control average (and standard deviation) run length performance of the ``egg'' parts for the missing strut defect, estimated from 5000 replications. For the TDA-based SPC method, both 0-dimensional and 1-dimensional features are individually and jointly monitored, and for the spectral SPC method, the first 15 LB eigenvalues are used. $m_0=200, \alpha=0.05$ for both methods.}
\label{TDAvsSpectral_AL_OC_MS}
\end{table}\vspace{1cm}

\section{Conclusions}
In this paper, we advanced for the first time the application of Topological Data Analysis (TDA) tools for the Statistical Process Control. The main application is the monitoring of complex, lattice-like parts, typically created via additive manufacturing. We developed a new permutation-based SPC technique based on the persistent homology of parts. The new TDA-based SPC method exhibits high sensitivity to typical AM defects, particularly using 0-dimensional and 1-dimensional features. We further conducted extensive simulations validating the chart’s average run length performance across various lattice-like objects, confirming its potential efficacy in industrial applications. In the comparison study between the proposed TDA-based method and the spectral method by \cite{zhaoEDC_Tech}, under small to moderate noise conditions, the proposed TDA-based SPC method demonstrates enhanced detection speed of both defect types in comparison to the spectral SPC method, validating its strength and efficiency for monitoring topological features in intricate AM structures, while under large noise conditions, the spectral method shows more consistent run-length performance against the proposed TDA-based method, potentially due to the additional edge information used in the FEM computation of the Laplace-Beltrami spectrum in the spectral method.

For future research, one should conduct a comprehensive demonstration of the proposed Phase II SPC scheme using a sequence of real-world, additively manufactured lattice-structured parts to validate its practical applicability and robustness in industrial settings. Furthermore, investigating the monitoring of 2-dimensional topological features, which were not addressed in this study, could enhance the sensitivity of the SPC chart to a broader range of defects in complex geometries, such as lattice-structured parts. Lastly, this paper focused on mesh data. To be able to measure and inspect internal features of complex parts, industrial tomographic machines are used, and therefore, voxel data needs to be considered. In recent work, \cite{zhaoEDC_PE} developed LB-spectrum techniques for voxel data, rather than meshes. Similarly, it is of interest to extend the TDA-based SPC methods from point cloud data to voxel-type data obtained from tomography machines.
    
\subsection*{Data Availability Statement.} The supplementary materials contain computer code that implements all methods discussed and generates all the simulated datasets used in the examples. The supplementary materials are available to download at: \url{https://drive.google.com/drive/folders/1yz9N3H_M7iiEAk2uMXwYWD4qmdg7qk-1}.

\subsection*{Disclosure of interest.} The authors have no conflicts of interest to report.

\subsection*{Acknowledgments.} This work was funded by NSF grant CMMI 2121625. We thank Professors Bianca Colosimo and Marco Grasso (Politecnico di Milano, Italy) for allowing us use their ``egg" datasets.

\appendix
\section{Concepts on Topological Data Analysis}
\label{app:A}
In this section, we review some introductory notions and concepts about Topological Data Analysis (TDA). They are the basis of the persistent homology methods utilized in this paper. In TDA, we observe a set of points in $\mathbb{R}^n$, i.e., a point cloud,  which topologically does not have in itself any interesting property. It is assumed that these points are measurements on some manifold, and the idea of TDA is to make statistical inferences on the underlying manifold shape from the point cloud data. To do this in practice, a triangulation is introduced in the point cloud, as discussed below.

\begin{figure}[h]
\centering
\includegraphics[width=12cm,height=4cm]{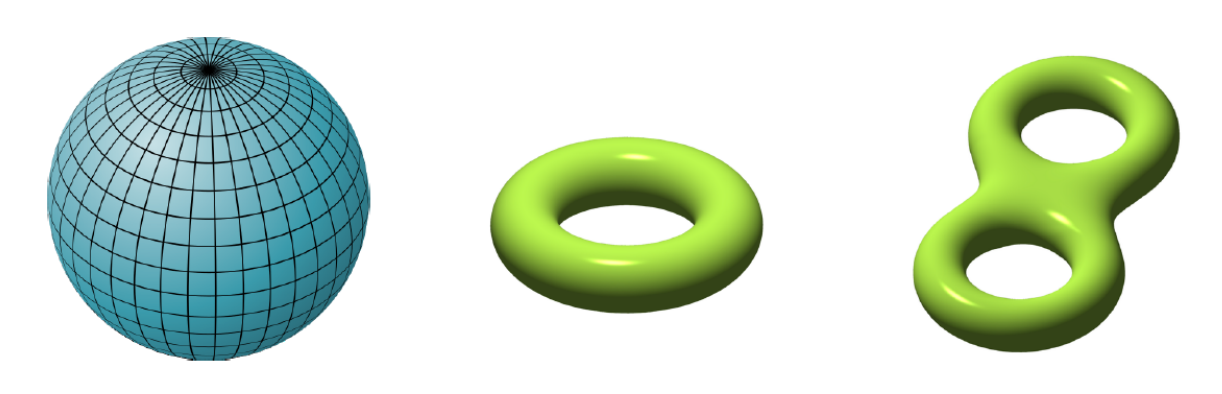}
\caption{The sphere has connectivity 0, the torus 2, and the ``pretzel'' (double torus), 4. This is the largest number of closed cuts that can be made without dissecting a piece from the surface.}
\label{fig: SphereTorus}
\end{figure}

Intuitively, every transformation of a geometric figure in which relations of {\bf \em adjacency} of its various parts are not destroyed is called continuous; if the adjacencies are not only not destroyed, but no new ones arise, then the transformation is called {\bf \em topological}. Topological transformations are therefore not only continuous but are one-to-one (single valued and continuous both ways) and are commonly described as if the objects were made of elastic ``rubber'' that can be deformed as desired without breaking. Thus, under a topological transformation, parts of an object that are in contact remain in contact, and parts not in contact remain not in contact; neither breaks nor fusions can arise.  {\bf \em Topological properties} of an object are those that are invariant with respect to topological transformations, and {\bf \em Topology} is the study of the topological properties of objects. For instance, as shown in the left and middle panels of Figure \ref{fig: SphereTorus}, the torus and the sphere are {\em topologically distinct} objects; one cannot be transformed into the other by a topological transformation. In contrast, an ellipsoid and a sphere are topologically equal; one can be transformed topologically to the other, and we say they are {\bf \em homeomorphic}. Topological properties are therefore {\em intrinsic}, as they do not refer to the ambient space where the objects exist. In the torus, shown in the middle panel of Figure \ref{fig: SphereTorus}, up to two closed cuts can be made without dissecting a piece, while on an 8-shaped ``pretzel'', shown in the right panel of Figure \ref{fig: SphereTorus}, up to four cuts along closed curves can be made without dissecting a piece, and so on. The largest number of closed curves on an object such that they do not dissect a part is called the {\bf \em order of connectivity}. The order of connectivity is a topological invariant.

{\bf \em Combinatorial Topology} considers topological properties of any triangulation of a surface, and hence, a property of the surface itself. Combinatorial methods are not only applicable to surfaces of intrinsic dimension 2 but also to manifolds of higher dimensions. In a 3D manifold, triangulations are made of tetrahedra (or 3D simplices).

\subsection{Topological Spaces and Simplicial Complexes} 
\begin{definition}
A {\bf \em Topological Space} is a set $R$ composed composed of elements called {\em points} in which certain subsets $A \subseteq R$, called {\em open sets} of $R$ satisfy the following two axioms:
\begin{enumerate}
\item the union of any number and the intersection of any finite number of open sets is an open set
\item the whole set $R$ and the empty set are open.
\end{enumerate}
\end{definition}

\begin{definition}
A topological space such that each point has a neighborhood homeomorphic to Euclidean $n$-dimensional space is called an {\bf \em $n$-manifold}. 
\end{definition} 

\begin{definition}
The {\bf \em genus} of an orientable surface is one-half its connectivity.
\end{definition}

\begin{definition}
The set of points in the $n-dimensional$ plane generated by a set of $n+1$ linearly independent points $e_0, e_1,...,e_n$ in $\mathbb{R}^m$, $0 \leq n \leq m$,  is called a {\bf \em $n$-simplex}  $T$ with vertices $e_0, e_1,...,e_n$, or $T=(e_0, e_1,...,e_n)$. The number of vertices minus one is the {\em dimension} of the simplex, usually denoted with a superscript, hence $T^n$ denotes an $n$-simplex. If the vertices in a simplex $T_1$ are a subset of the vertices in a simplex $T_2$, $T_1$ is said to be a {\em face} of $T_2$. $T_1^r$ is a {\em proper face} of $T_2^n$ if $0< r < n$ (the vertices and the whole $n$-simplex are not proper faces). For $n=3$, 0-faces are the vertices, 1-faces are the edges, 3-faces are triangles, and the 4-face tetrahedron is the 4-dimensional simplex itself. 
\end{definition}

\begin{definition}
A {\bf \em simplicial complex} $K$ is a set of simplices such that:
\begin{enumerate}
\item if $T_1$ is a simplex in $K$ and $T_2$ is a proper face of $T_1$, then $T_2$ is also in $K$;
\item the intersection of two simplices in $K$ is either a simplex in $K$ or is empty.
The set $K$ may be partially ordered as follows: $T_1 \in K$ precedes $T_2 \in K$ $(T_1 < T_2)$, if $T_1$ is a proper face of $T_2$.
\end{enumerate}
\end{definition}

\begin{figure}[h]
\centering
\begin{tabular}{cc}
\includegraphics[width=5cm,height=4cm]{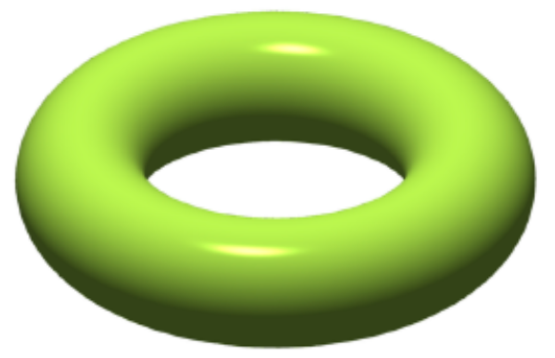}&\includegraphics[width=5cm,height=4cm]{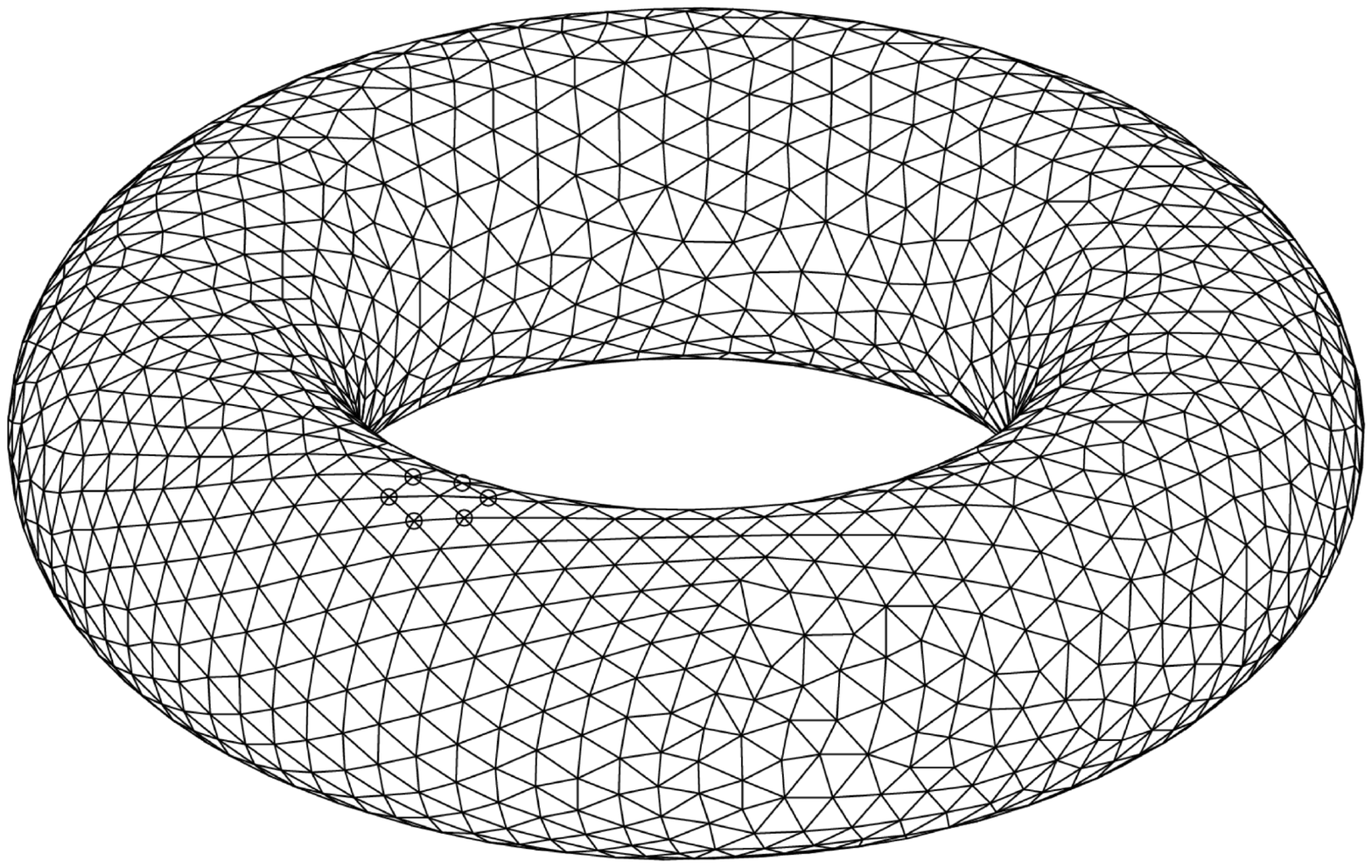}
\end{tabular}
\caption{A topological space $X$ and its triangulation $\hat P$ are homeomorphic.}
\label{fig:triangulation}
\end{figure}

\begin{definition}
A {\bf \em triangulation} $\hat P$ of a set of points $P$ in a topological space $X$ is a simplicial complex that covers the convex hull of the points in $P$ such that $\hat P$ and $X$ are homeomorphic (see Figure \ref{fig:triangulation}). 
\end{definition}

\subsection{Homology Group and Betti Numbers}

Topologists started working on the concept of Homology from the intuition that objects of similar shape have the same number of holes. Early authors tried to provide a classification of different manifolds, with the earliest such attempt being Euler's characteristic for regular polyhedra. \textbf{Homology}, a similar modern classification,  quantifies the number of holes or cycles in a manifold,  and is computed using simplicial complexes. 

\begin{figure}[H]
\centering
\includegraphics[width=14cm,height=3cm]{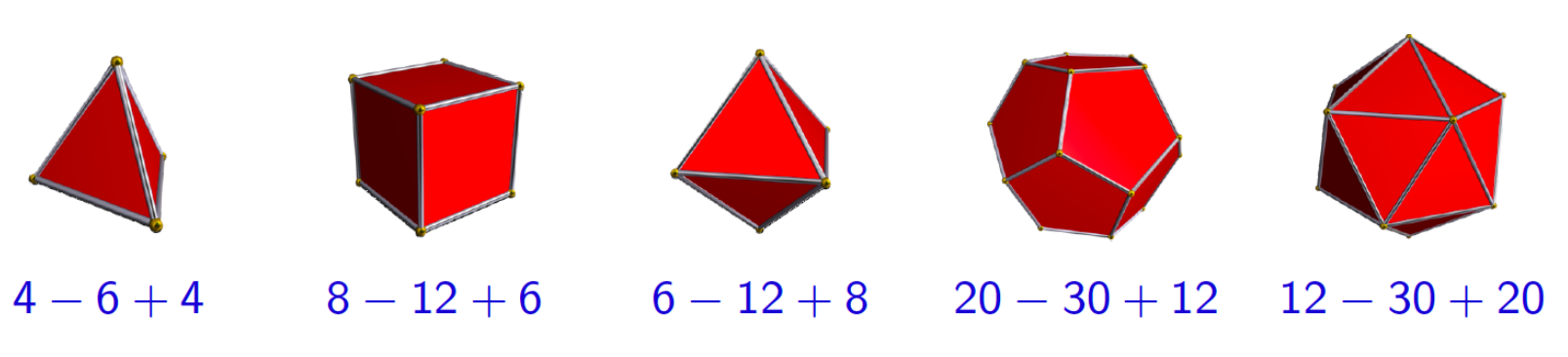}
\caption{Euler characteristics for the regular polyhedra $\chi = \alpha_0 - \alpha_1 + \alpha_2$. From \cite{Abdelkader}.}
\label{fig:Euler}
\end{figure}

\begin{definition}
Suppose that a planar complex $K$ consists of $\alpha_0$ points or vertices $(T^0)$, $\alpha_1$ edges ($T^1$), and $\alpha_2$ triangles ($T^2$). Then its {\bf \em Euler characteristic} (see Figure \ref{fig:Euler}) is defined as:
\[ \chi = \alpha_0 - \alpha_1 + \alpha_2 \]
\end{definition}
For a 3D complex, the Euler characteristic is:
\[ \chi = \alpha_0 - \alpha_1 + \alpha_2  - \alpha_3\]
where $\alpha_3$ is the number of tetrahedra ($T^3)$. For dimensions greater than 3, a triangulation contains up to $n$-dimensional simplices as elements, and its Euler characteristic is:
\[ \chi = \sum_{i=0}^n (-1)^i \alpha_i\]

All manifolds homeomorphic to a simplicial complex have the same Euler characteristic, since it is a topological invariant.

\begin{definition}
The {\bf \em Betti numbers} $\beta_p$ of a topological space $X$ measure the connectivity of the space utilizing a simplicial complex. In a simplicial complex on $X$, $\beta_0$ counts the number of connected components; $\beta_1$ counts the number of one-dimensional holes, $\beta_2$ counts the number of two-dimensional holes (voids), etc.
\end{definition}

Since, in practice, the Betti numbers of a manifold are computed using incidence matrices based on a simplicial complex, the complex must first be created. One way to do this is with the so called {\bf \em \v{C}ech complex}, originally due to \cite{Alexandrov1}, defined as the union of balls of diameter $\epsilon$ centered at each vertex such that the intersection of the balls is not empty. 

Computationally, the Vietoris-Rips (VR) complex is the most efficient as it only requires the computation of pairwise distances between all vertices.

\begin{figure}[h]
\centering
\includegraphics[width=8cm,height=4cm]{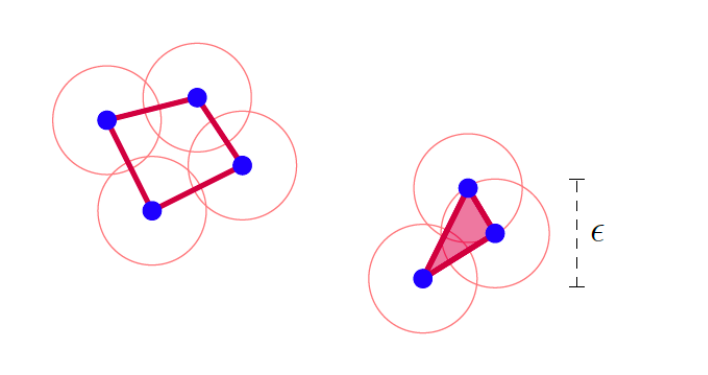}
\includegraphics[width=5cm,height=4cm]{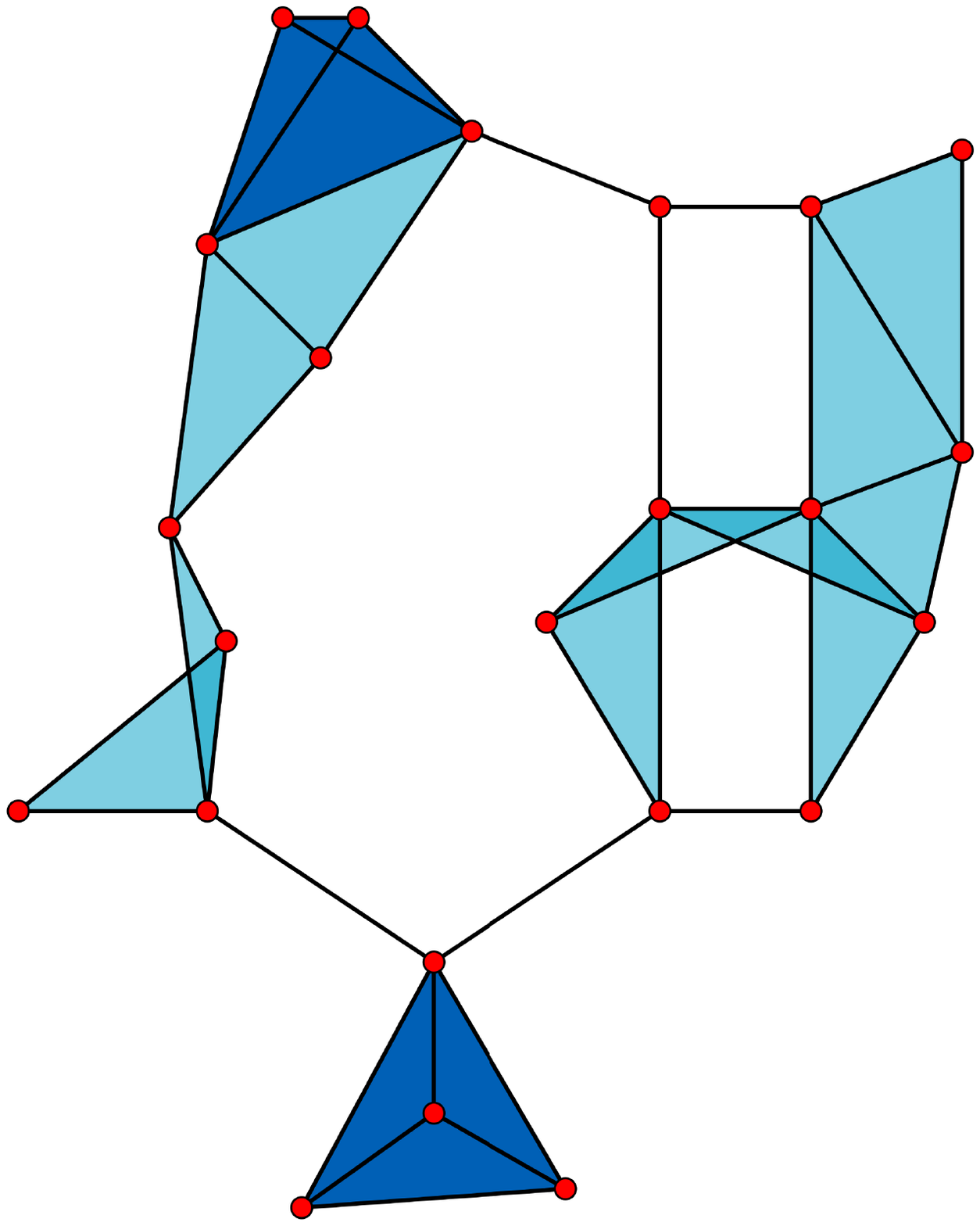}
\caption{Left: the Vietoris-Rips complex of seven points for a given value of $\epsilon$.  For this complex, $\beta_0=2$ (two components), $\beta_1=1$ (one hole) and $\beta_k=0$ for $k >1$. From \cite{Wright}. Note that the circles are for visualization purposes and are not part of the complex. The complex has seven 0-simplices, seven 1-simplices, and one 2-simplex, shaded in pink. Right: the Vietoris-Rips complex of 23 points (from \cite{WikiVietorisRips}). This complex has 0-simplices (red vertices), 1-simplices (edges), 2-simplices (light blue triangles), and 3-simplices (dark blue tetrahedra).}
\label{fig:VR}
\end{figure}

\begin{definition}
Given a set of points $P$ from some manifold, the {\bf \em Vietoris-Rips complex} $V(P,\epsilon)$ consists of the simplices with vertices in $P$ and diameter at most $2\epsilon$. That is, a simplex $\sigma$ is included in the complex if each pair of vertices in $\sigma$ is at most $2\epsilon$ apart (see Figure \ref{fig:VR}).
\end{definition}

\subsection{Persistent Homology}
In order to compute the homology of a manifold from a point cloud sampled from it, it is necessary to create a simplicial complex, either a \v{C}eck complex or a Vietoris-Rips complex (notice that to create a simplicial complex with either of these methods, we are introducing a metric into the topological space).  In either case, the difficulty is what value of $\epsilon$ to choose, since different values of $\epsilon$ will result in different topological properties. A solution to this problem is to {\em vary} $\epsilon$ and compute the corresponding homology. The homological properties that are {\bf \em persistent} over time as we vary $\epsilon$ provide a description of the topology of the underlying manifold. In TDA, 0-degree and 1-degree homologies are most frequently used due to computational constraints.

As $\epsilon$ is increased and a VR complex is created from points in a point cloud $P$, it will result in an ordered sequence, or {\bf \em filtration}, of complexes, where complexes computed at two adjacent values of $\epsilon$ satisfy $R_k \subset R_{k+1}$. That is, increasing $\epsilon$ can only add more elements to the complex. For the Vietoris-Rips complex, the filtration can be represented as
\begin{equation}
R_0 \hookrightarrow R_1 \hookrightarrow R_2 \hookrightarrow ... R_{max}
\label{eq:filtration}
\end{equation}
where $R_0 = V(P,0)$ and $R_{max}=VR(P,d_P)$ where $d_P$ is the maximum diameter of the point cloud. 

\begin{figure}
\centering
\includegraphics[width=6.5cm,height=3.75cm]{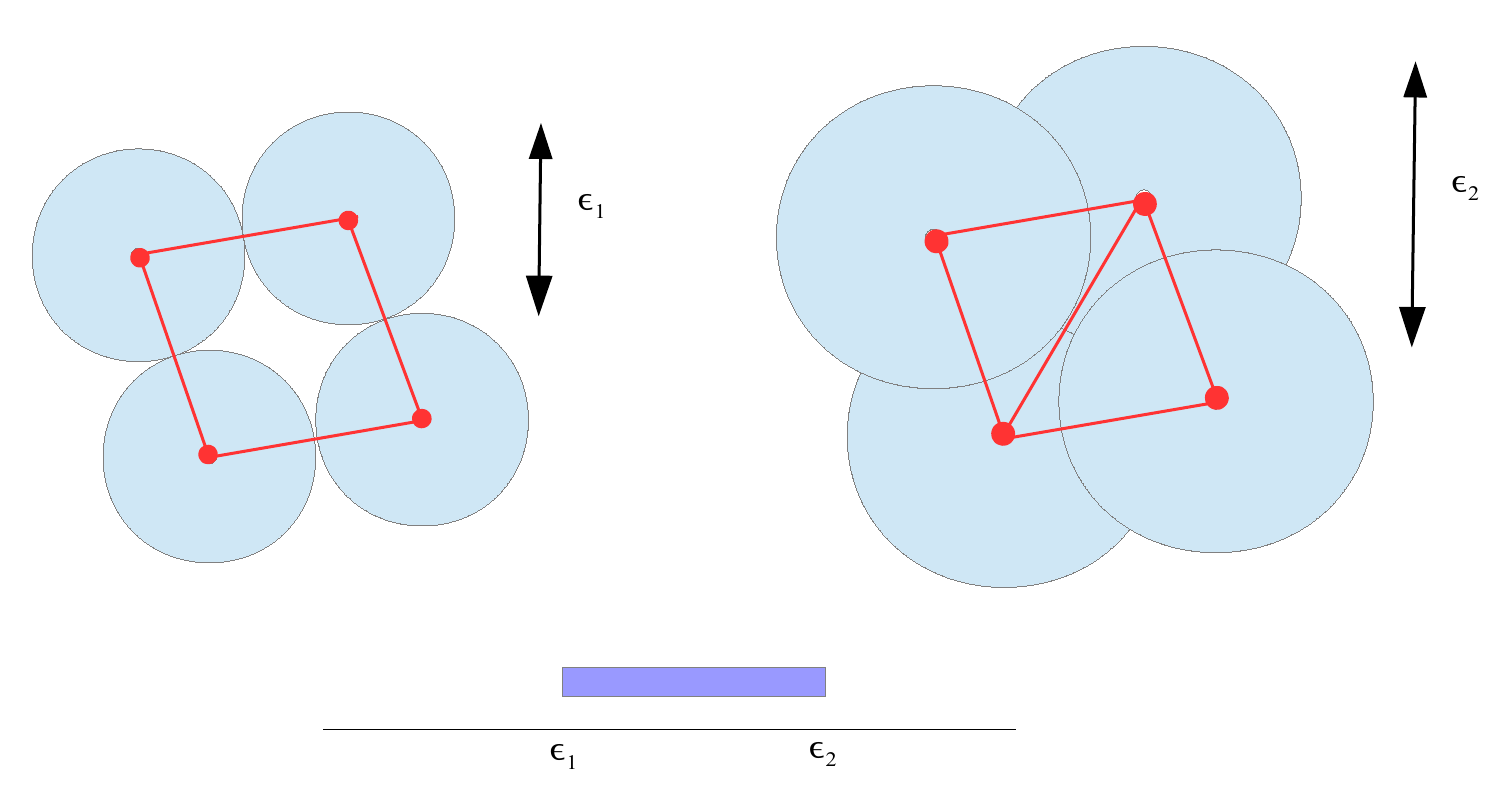} \includegraphics[width=7.5cm,height=3.75cm]{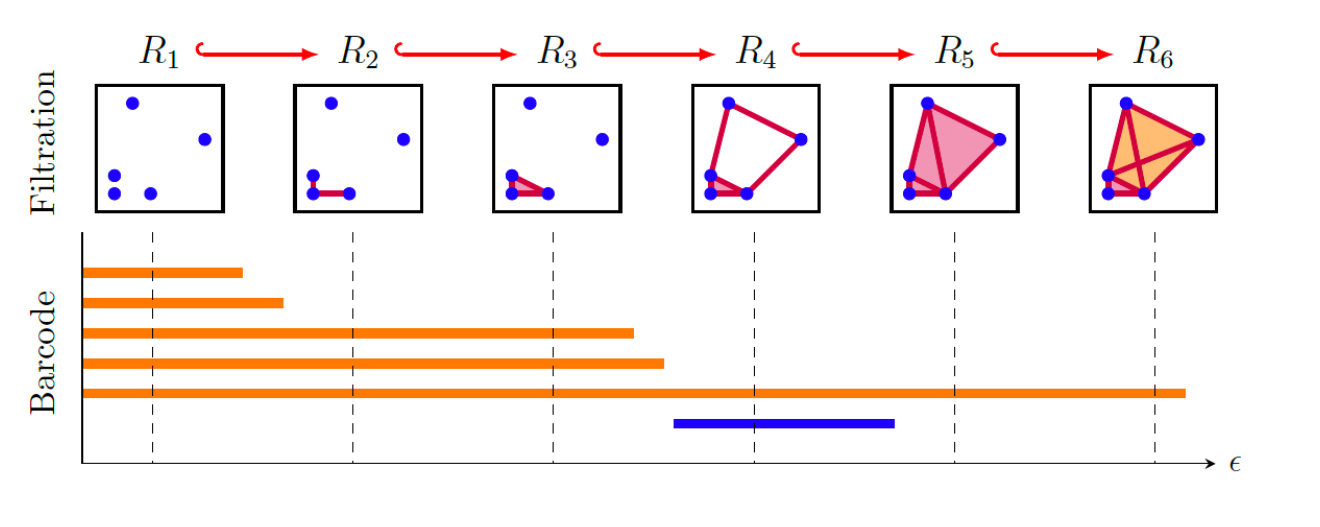}
\caption{Left: a VR complex at two values of $\epsilon$. At $\epsilon_1$, only the 1-simplices are within a distance $2\epsilon_1$, and a ``hole'' (1-cycle) exists. At $\epsilon_2 > \epsilon_1$, two diagonally opposite vertices are also within distance $2\epsilon_2$, and hence a new 1-simplex is introduced, which ``kills'' the hole. The hole persisted from $\epsilon_1$ to $\epsilon_2$, and this would generate a ``bar'' in a barcode diagram as shown on the left, bottom. Note the relation between the two complexes, with $R_1 \subset R_2$. On the right: a sequence of VR complexes as a function of $\epsilon$ and the corresponding barcode. The orange bars denote connected components (homologies of order 0) while the blue bar denotes a hole (homology of order 1). As $\epsilon$ increases, the hole is born and dies soon after. From \cite{Wright}.}
\label{fig:persistence}
\end{figure}

If we imagine that $\epsilon$ is varied according to some ``clock'', we say that a homology class $[c]$ represented by a $p-cycle$ is {\bf \em born} at $R_s$ if $[c]$ is not supported in $R_r$ for $r<s$ but exists in for $r=s$, and that it {\bf \em dies} at $R_t$ if $t$ is the smallest index at which $[c]$ is present. The interval of ``time'' $t-s$ is how long the cycle $[c]$ {\em persisted}, and it provides the {\bf \em persistent homology} of the point cloud. A {\bf \em persistence diagram} is obtained by plotting each pair $(s,t)$ on the plane. Alternatively, a {\bf \em barcode plot} can be plotted by adding an interval to the pair (see Figure \ref{fig:persistence}).

\end{document}